\documentclass[aps,prb,groupedaddress,twocolumn,superscriptaddress, longbibliography]{revtex4-1}
\usepackage{enumitem}
\usepackage{paralist}

\usepackage{graphicx} 
\usepackage{bpchem}
\usepackage{dcolumn}
\usepackage{bm}
\usepackage{amsmath,amsthm,amssymb}
\usepackage{color}
\usepackage{hyperref}
\usepackage{multirow}
\usepackage{array}
\nocite{TitlesOn}
\usepackage{xr}

\newcommand{\cuse}{Cu$_2$OSeO$_3$}

\usepackage{hhline}

\begin{document}
\title{New magnetic phase of the chiral skyrmion material  \cuse}
\author{F. Qian}
\altaffiliation{Nanjing University of Aeronautics and Astronautics, China}
\author{L.J. Bannenberg}
\affiliation{Faculty of Applied Sciences, Delft University of Technology, Mekelweg 15, 2629 JB Delft, The Netherlands}
\author{H. Wilhelm}
\affiliation{Diamond Light Source Ltd., Chilton, Didcot, Oxfordshire, OX11 0DE, United Kingdom}
\author{G. Chaboussant}
\affiliation{Laboratoire L\'{e}on Brillouin, CEA-Saclay, 91191 Gif sur Yvette, France}
\author{L. M. Debeer-Schmitt}
\affiliation{Oak Ridge National Laboratory, Oak Ridge, Tennessee 37831, USA}
\author{M. P. Schmidt}
\affiliation{Max Planck Institute for Chemical Physics of Solids, N{\"o}thnitzer-Stra\ss e 40, 01187 Dresden, Germany}
\author{A. Aqeel  }
\altaffiliation[Present Address: ]{University of Regensburg, Germany}
\affiliation{Zernike Institute for Advanced Materials, University of Groningen, Nijenborgh 4, 9747 AG Groningen, The Netherlands}
\author{T.T.M. Palstra}
\altaffiliation [Present Address: ] {University of Twente, The Netherlands}
\affiliation{Zernike Institute for Advanced Materials, University of Groningen, Nijenborgh 4, 9747 AG Groningen, The Netherlands}
\author{E. Br\"uck}
\author{A.J.E. Lefering}
\author{C. Pappas}
\affiliation{Faculty of Applied Sciences, Delft University of Technology, Mekelweg 15, 2629 JB Delft, The Netherlands}
\author{M. Mostovoy}
\affiliation{Zernike Institute for Advanced Materials, University of Groningen, Nijenborgh 4, 9747 AG Groningen, The Netherlands}
\author{A. O. Leonov}
\affiliation{Chiral Research Center, Hiroshima University, Kagamiyma, Higashi Hiroshima, Hiroshima 739-8526, Japan}
\affiliation{Department of Chemistry, Faculty of Science, Hiroshima University, Kagamiyma, Higashi Hiroshima 739-8526, Japan}



\begin{abstract}
The lack of inversion symmetry in the crystal lattice of  magnetic materials gives rise  to complex non-collinear spin orders through interactions of relativistic nature, resulting in interesting physical phenomena, such as emergent electromagnetism. Studies of cubic chiral magnets revealed a universal magnetic phase diagram, composed of helical spiral, conical spiral and skyrmion crystal phases. Here, we report a remarkable deviation from this universal behavior. 
By combining neutron diffraction with magnetization measurements we  observe a new multi-domain state in 
\cuse.  
Just below the upper critical field at which the conical spiral state disappears, the spiral wave vector rotates away from the magnetic field direction. 
This transition gives rise to large magnetic fluctuations. 
We clarify the physical origin of the new state and discuss its multiferroic properties.

\end{abstract}
\maketitle


\section*{Introduction}

Chiral magnets show a variety of periodically modulated spin states --  spirals\cite{Dzyaloshinskii:1964JETP,Bak:1980ud}, triangular and square arrays of skyrmion tubes\cite{Bogdanov_1999JMMM,Muhlbauer:2009bc,yu2010real,Nayak:2017hv,Karube2016,Nakajima:2017jx}, and  a cubic lattice of monopoles and anti-monopoles\cite{kanazawa2012possible,
kanazawa2016critical} --  which can be viewed as magnetic crystals of different symmetries and dimensionalities.
%
%
These competing magnetic superstructures show high sensitivity to external perturbations, allowing for the control of phase boundaries  with  applied  electric fields and stresses
\cite{okamura2016transition,nii2015uniaxial}.
%
The non-trivial topology of  multiply-periodic magnetic states gives rise to emergent electromagnetic fields and unconventional spin, charge, and heat transport
\cite{Lee:2009gp,neubauer2009topological,Zang:2011vb,Schulz:2012NatPh,Nagaosa:2013cc}.
%
The stability and small size of magnetic skyrmions as well as low spin currents required to set them into motion paved the way to new prototype memory devices\cite{Fert:2013fq,jiang2015blowing,moreau2016additive,woo2015observation,Fert:2017hv}.  

Recent studies of chiral cubic materials hosting skyrmions, such as the itinerant magnets, MnSi and FeGe, and the Mott insulator, Cu$_2$OSeO$_3$, showed that they exhibit the same set of magnetic states with one or more long-period spin  modulations and undergo similar transitions under an applied magnetic field \cite{Bauer:2016fd}. 
This universality is a result of non-centrosymmetric cubic lattice symmetry and  the hierarchy of energy scales \cite{Bak:1980ud,1Nakanishi:1980SS,Belitz:2006vd}. 
The transition temperature $T_c$ is determined by the ferromagnetic (FM) exchange interaction $J$. 
The relatively weak antisymmetric Dzyaloshinskii-Moriya (DM) interaction  with the  strength $D$ proportional to the spin-orbit coupling constant $\lambda$, renders the uniform FM state unstable towards   a helical spiral modulation  \cite{Dzyaloshinskii:1964JETP,Moriya:1960go}.  
It determines the magnitude of the modulation wave vector ${\bm Q}$  and the value of the critical field $H_{C2}$, above which the spiral modulation is suppressed. 
In contrast to low-symmetry systems \cite{Togawa_2012PRL, Kezsmarki_2015NatMa}, the DM interaction in cubic chiral magnets does not impose constraints on the direction of the spiral wave vector\cite{Bak:1980ud}.
%
The  direction of the wave vector is controlled by the applied magnetic field and magnetic anisotropies of higher order in $\lambda$. 
In the helical spiral phase observed at low magnetic fields, magnetic anisotropies pin the direction of $\bm Q$  either along one of the cubic body diagonals, as in MnSi, or along the cubic axes, as in FeGe or Cu$_2$OSeO$_3$. 
%
The competition between the Zeeman and  magnetic anisotropy energies sets the critical field $H_{C1}$ of the transition between the helical and conical spiral states, above which $\bm Q$ is parallel to the applied magnetic field. 
%
In the multiply-periodic skyrmion crystal state, the spiral wave vectors are perpendicular to the field direction, which is favored by the non-linear interaction between the three helical spirals. 
\begin{figure*}
\includegraphics[width= 1\textwidth]{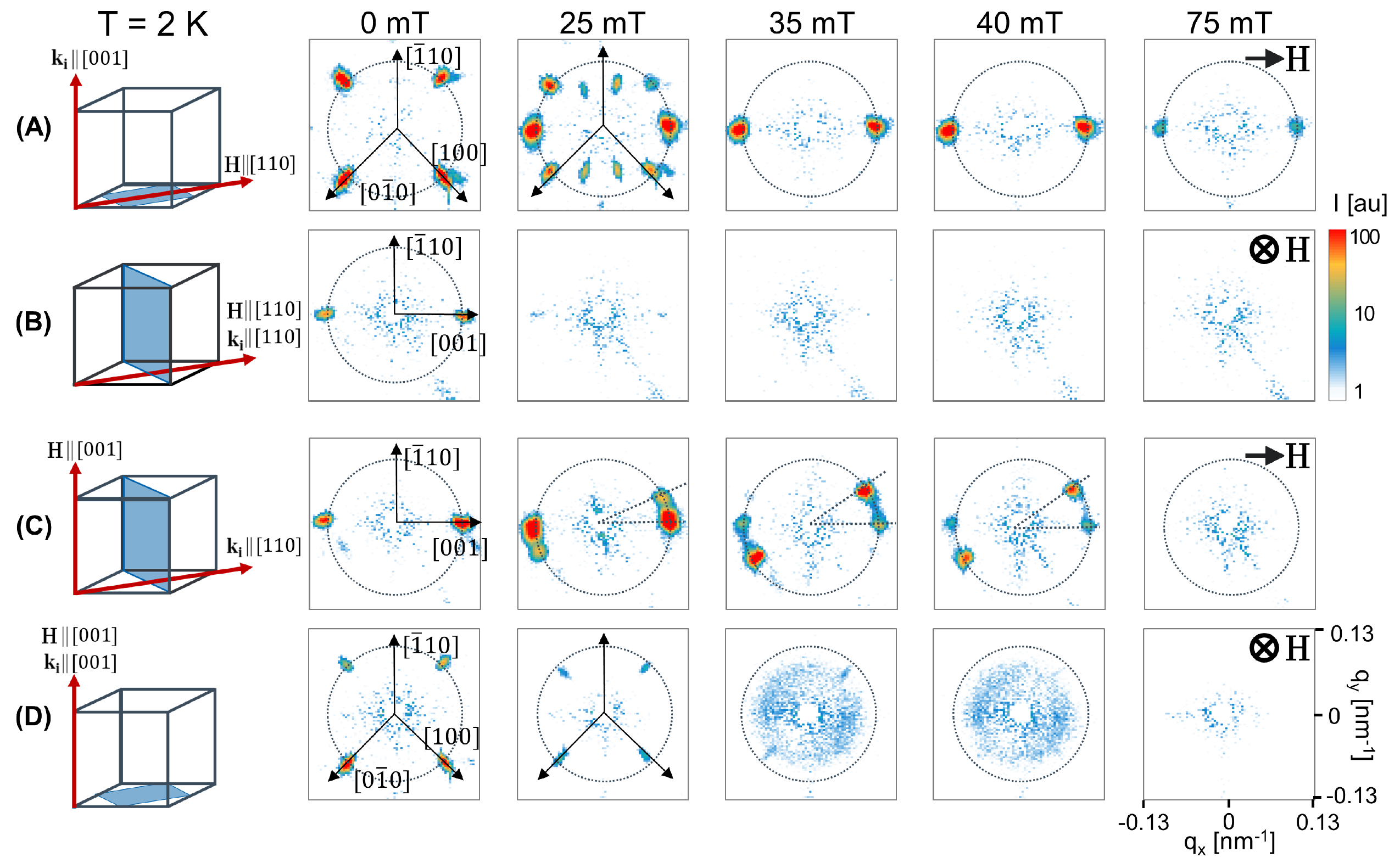}
\caption{\textbf{Magnetic field dependence of the \cuse\ SANS patterns at $\mathbf{T=2}$~K}. The first column illustrates the geometry of the experiment:  $\bm H\| [110]$ in {\bf A}, {\bf B} and   $\bm H\| [ 001 ]$ in {\bf C}, {\bf D}. The orientation of the neutron beam wavevector is: $\bm k_i \perp \bm H$ in  {\bf A}, {\bf C}  and $\bm k_i \| \bm H$ in {\bf B}, {\bf D}. The blue planes illustrate the SANS detection plane. The radius of the dashed circles on the SANS patterns corresponds to the modulus of the helical spiral propagation vector ${Q}=2\pi/\ell\sim{0.1}\ \text{nm}^{{-1}}$, with $\ell\sim60$~nm being the  pitch of the helix. } 
\label{Fig:SANS}
\end{figure*}

Here, we report a remarkable deviation from this well established universal behavior.  
By Small Angle Neutron Scattering (SANS) and magnetic measurements we observe a new low-temperature magnetic phase of \cuse. 
At relatively high magnetic fields, 
$\bm Q$ tilts  away from the magnetic field vector, $\bm H$, when this is directed along the $[001]$ crystallographic direction favored by anisotropy at zero field. 
%
This transition occurs where it is least expected -- at $H$ close to $H_{C2}$ where the dominant Zeeman interaction favors $\bm Q \| \bm H$  and at low temperatures where thermal spin fluctuations that can affect the orientation of $\bm Q$ are suppressed. 
%
The re-orientation of the spiral wave vector is accompanied by strong diffuse scattering, reminiscent of the pressure-induced partially ordered magnetic state in MnSi.  
%
The instability of the conical spiral state at high applied magnetic fields  can be considered as a re-entrance into the helical state, although $\bm Q$  in the ``tilted conical spiral'' state is not close to high-symmetry points.
The new phase of \cuse~ is sensitive to the direction of the applied magnetic field: for  $\bm H  \| \langle110 \rangle$  no tilted spiral state is observed.
Instead, we find that the helical-to-conical spiral transition splits into two transitions occurring at slightly different magnetic fields.

We show theoretically that the tilted spiral state originates from the interplay of competing anisotropic spin interactions, which is generic to chiral magnets and  may be important for understanding  the structure of metastable skyrmion crystal states\cite{Karube2016,Nakajima:2017jx, Bannenberg:2016kh}. 
This interplay is particularly strong in \cuse\ due to the composite nature of spin of the magnetic building blocks \cite{Janson:2014uo}.
The transition to the new state in multiferroic \cuse~ should have a strong effect on the magnetically-induced electric polarization. It should also affect the spin-Hall magnetoresistance\cite{Aqeel2016} and modify the spin-wave spectrum. 

\begin{figure*}
\includegraphics[width= 1\textwidth]{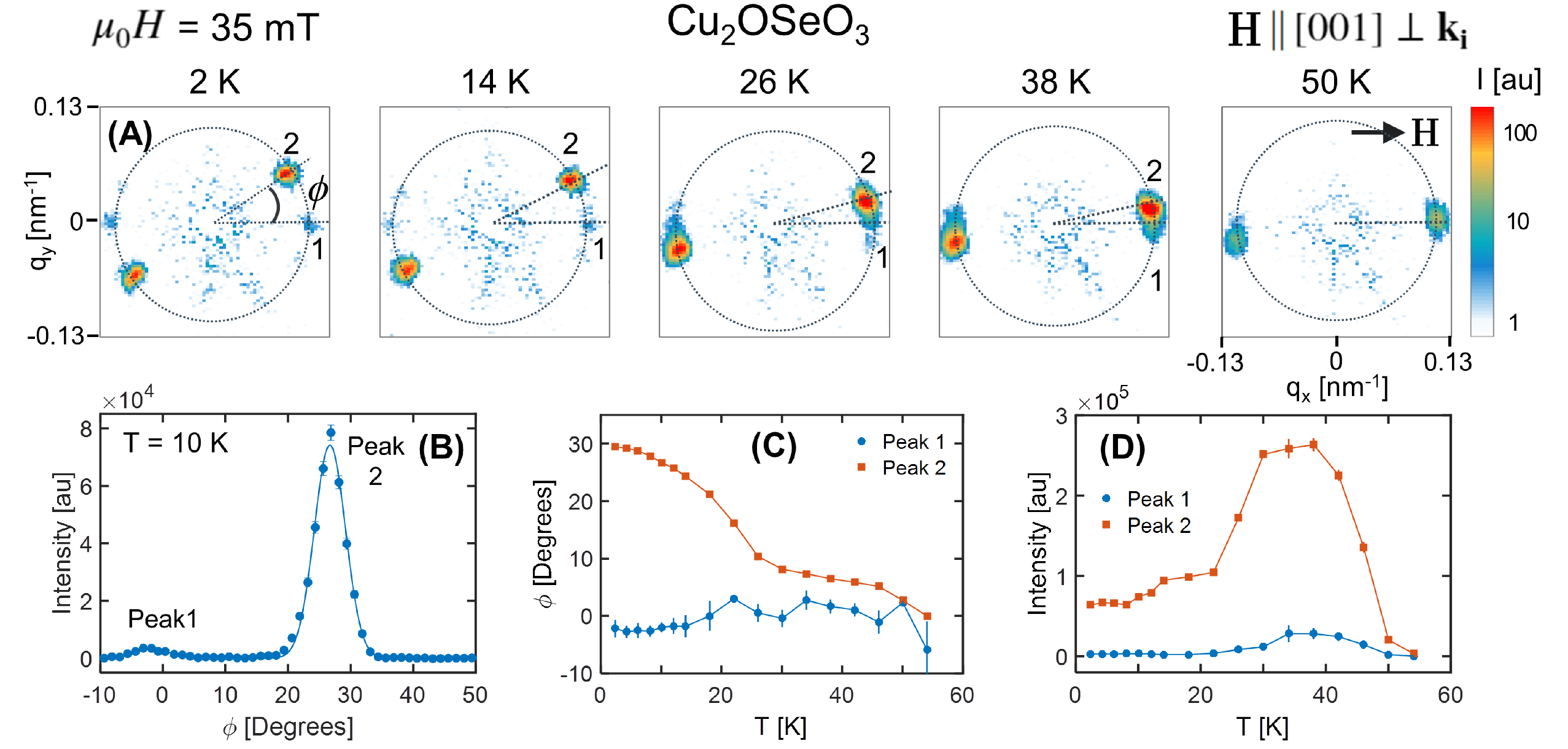}
\caption{\textbf{Temperature dependence of the tilt}. ({\bf A}) Temperature dependence of the SANS patterns and  ({\bf B}-{\bf D}) the corresponding data analysis. The  dashed circles on the SANS patterns  are a guide to the eye and have a radius of  $Q=2\pi/\ell \sim 0.1\ \text{nm}^{-1}$.  The angular dependence of the scattered intensity along the circle with radius $Q$ is given in {\bf B} for  $T=10$~K. The solid lines indicate a fit with two Gaussian peaks labeled  $1$ and $2$ in the SANS patterns. The temperature dependence of the angular positions and integrated intensities of the two peaks is shown  in {\bf C} and {\bf D} respectively.} 
\label{Fig:SANSvarT}
\end{figure*}

\section*{Results}

First hints for the existence of the new phase came from the anomalous field dependence of the magnetization, $M$, and the ac magnetic susceptibilities, $\chi'$ and $\chi''$, shown in the Supplementary Sec.~\ref{sec:susc}. A direct confirmation was provided by SANS, which probes correlations perpendicular to the incoming neutron beam wavevector $\bm k_i$.
For this reason, we performed our measurements in two crystallographic orientations and for each orientation in two complementary configurations, $\bm H \bot \bm k_i $ and $\bm H \| \bm k_i$, thus providing a full picture of the effect of the magnetic field on the magnetic correlations. 

A selection of patterns obtained at $T={2}$~K is shown in  Fig.~\ref{Fig:SANS}A,B for $\bm H \| [ 110 ]$, and  Fig.~\ref{Fig:SANS}C,D for $\bm H \| [ 001 ] $.
At zero field the SANS patterns show four peaks along the diagonal directions in  Fig.~\ref{Fig:SANS}A,D for $\bm k_i  \|  [ 001 ]$, and two peaks along the horizontal axis in  Fig.~\ref{Fig:SANS}B,C for $ \bm k_i \|  [ 110 ]$. 
These are the magnetic Bragg peaks of the helical spiral state with  wave vectors along the three equivalent $\langle001\rangle$  crystallographic directions.

At $\mu_0 H=25$~mT, the scattered intensity vanishes for $\bm H \| \bm k_i $ (Fig.~\ref{Fig:SANS}B,D) due to the re-orientation of the spiral wave vector along the magnetic field at the transition to the conical spiral phase. 
On the other hand, for $\bm H \bot \bm k_i $ and $\bm H \|[ 110 ]$, Fig.~\ref{Fig:SANS}A shows the coexistence of  helical spiral and conical spiral peaks (additional weak peaks are attributed to multiple scattering). 
Thus  the helical-to-conical transition for $\bm H \| [ 110 ]$ is not a simple one-step process.
Re-orientation first occurs in the helical spiral domain with the wave vector perpendicular to the field direction, $\bm Q \| [001]$. It is  followed by a gradual re-orientation of the wave vectors of the other two helical spiral domains. 
Upon a further increase of the magnetic field, the conical spiral  peaks weaken in intensity and disappear at the transition to the field-polarized collinear spin state, which for $\bm H \| \langle 110 \rangle$  occurs above $75$~mT.

The unexpected behavior, signature of the new phase, is seen in the evolution of SANS patterns  for $\bm H \| [ 001 ]$  in Fig.~\ref{Fig:SANS}C and D. 
For $\bm H \bot \bm k_i $  (Fig.~\ref{Fig:SANS}C), the Bragg peaks broaden along the  circles with  radius  $Q$ and eventually split into two well-defined peaks at  $35$ and $40$~mT. 
This is surprising because, in this configuration the two peaks  along the horizontal axis  correspond to the spiral with the wave vector parallel to both the magnetic field and the cubic axis. 
Thus no re-orientation is  expected for the spiral domain favored by both the Zeeman interaction and magnetic anisotropy. 
In addition to the splitting of the Bragg peaks, in the complementary configuration of $\bm  H \| \bm k_i $ shown  in Fig.~\ref{Fig:SANS}D,  a broad ring of scattering develops well inside the circle with radius $Q$.


%
With increasing temperature the splitting of the Bragg peaks becomes smaller and  disappears at $\sim50$~K as shown in Fig.~\ref{Fig:SANSvarT}A. 
At $T=10$ K the scattered intensity on the circle with radius $Q$, when plotted against the  azimuthal angle $\phi$, consists of two Gaussians peaks, labeled $1$ and $2$ (Fig. \ref{Fig:SANSvarT}B). 
These are centered at two distinct angles, which vary with temperature and their difference reaches ${30}^\circ$ at $T=2$~K (Fig.~\ref{Fig:SANSvarT}C).  
The  integrated intensities  depicted in Fig.~\ref{Fig:SANSvarT}D show that  peak 2, which splits away from the conical spiral peak~1, is by far the more intense one. Its intensity goes through a maximum at $\sim$35 K and then decreases at low temperatures, possibly because the optimum Bragg condition is not fulfilled any longer as the peak moves away from the magnetic field direction.  

\begin{figure*}
\includegraphics[width= 1\textwidth]{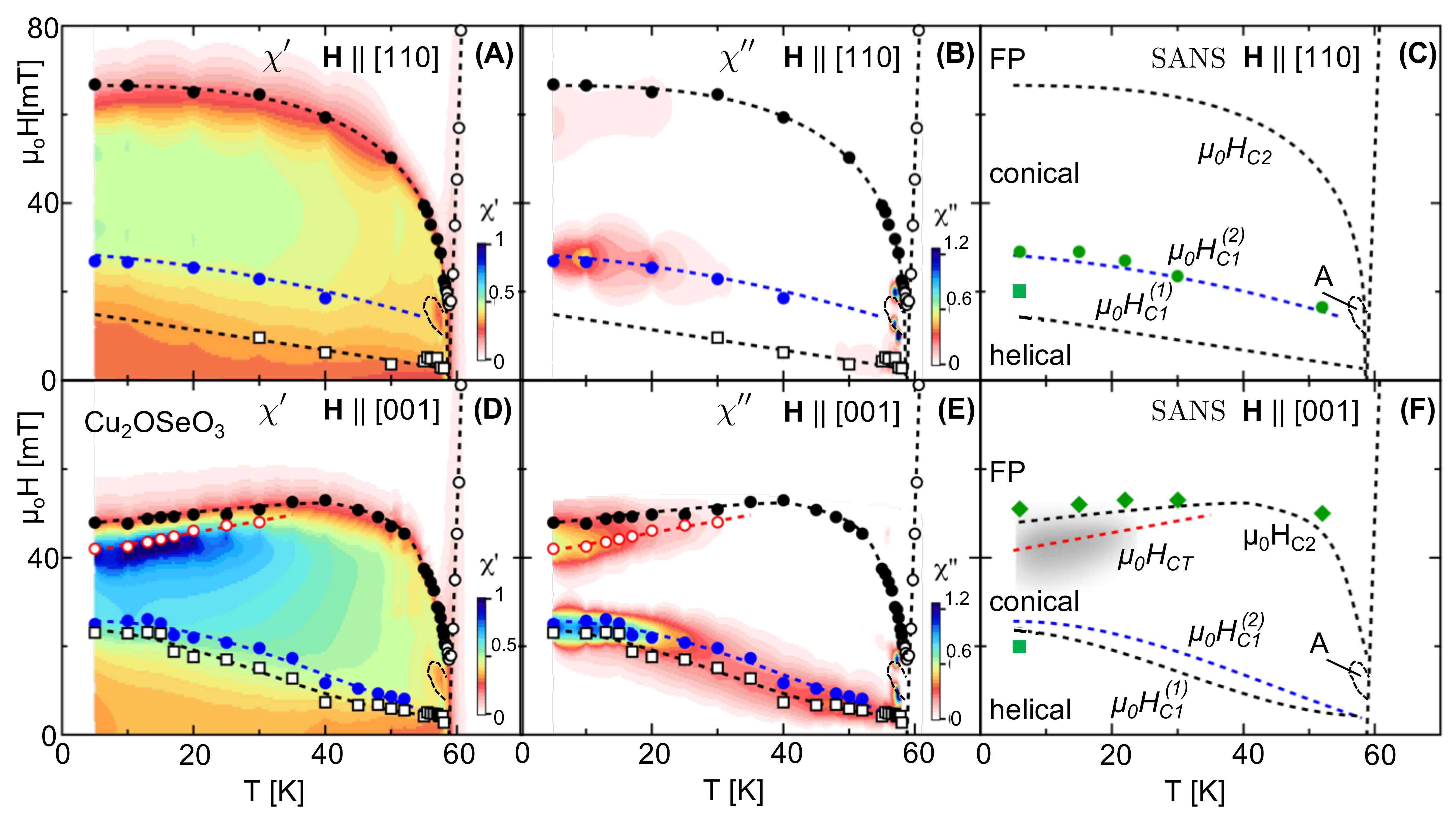}
\caption{\textbf{Phase diagrams from ac magnetic susceptibilities}. 
Contour plots of $\chi'$ and $\chi''$ at a frequency of 10~Hz and phase boundaries obtained by SANS for $\bm H \|[ 110 ] $, in {\bf A}-{\bf C}, and $\bm H \| [ 001]$, in {\bf D}-{\bf F}. 
The units for $\chi'$ and $\chi''$ are $10^{-4}$ and $10^{-6}$~m$^{3}$/mol$_\text{Cu}$, respectively. The helical, conical, $A$, tilted spiral (TS)  and field polarized (FP) phases are indicated in  {\bf C} and {\bf F}. 
The phase boundaries determined from the susceptibility are illustrated by the  symbols and the dashed lines  in {\bf A}, {\bf B}, {\bf D}, {\bf E}. 
They correspond to the peaks of $\chi''$, with the exception of $\mu_0 H_{C2}$, which is defined by the inflection point of $\chi'$ vs $\mu_0 H$. 
These criteria are the same as in our previous study\cite{Qian2016}. 
At low fields, two lines  $\mu_0 H_{C1}^{(1)}$ and $\mu_0 H_{C1}^{(2)}$ are identified below $50$~K. 
Just below $\mu_0 H_{C2}$, a red dashed line denoted as $\mu_0 H_{CT}$ in {\bf D}-{\bf F}, marks the onset of the ``tilted spiral'' state for  $\bm H \|\langle 001\rangle$. 
The phase boundaries determined from SANS are illustrated by the green symbols in  {\bf C} and {\bf F}. 
The shaded grey area just below $\mu_0 H_{C2}$ in {\bf F} marks the region where the ring of scattering emerges for $\bm H \|[ 001] \| \bm k_i $.  
}
\label{Fig:BT}
\end{figure*}

Our experimental findings are summarized in Fig.~\ref{Fig:BT}, which  shows contour plots of the real and imaginary  susceptibilities, $\chi'$ and $\chi''$,  as well as the phase boundaries obtained by SANS. 
Close to $T_c$, the transition from the helical to the conical phase is marked by a single $\mu_0 H_{C1}(T)$ line, which at low temperatures  evolves into two lines,  $\mu_0 H_{C1}^{(1)}(T)$ and $\mu_0 H_{C1}^{(2)}(T)$, derived from the two adjacent $\chi''$ peaks (see Fig.~\ref{Fig:ac}C,F and the discussion in the Supplement). 
The most prominent difference  between the two field orientations appears close to $\mu_0 H_{C2}$ below ${30}$~K. 
In this field and temperature range, clear maxima are seen in both $\chi'$ and $\chi''$ for $\bm H \| [ 001 ]$. 
These define a new line $\mu_0 H_{CT}(T)$ (red dashed line in Fig.~\ref{Fig:BT} D-F), which shifts slightly to lower magnetic fields with decreasing temperature.

The boundaries determined from the SANS measurements, shown  in Fig.~\ref{Fig:BT}C,F, are in excellent agreement with those derived from  susceptibility. 
Furthermore, it is remarkable that the shaded area in Fig.~\ref{Fig:BT}F, which marks the region, where the ring of scattering shown in Fig.~\ref{Fig:SANS}D emerges for  $\bm H \|\langle 001\rangle \| \bm k_i $, coincides with the maxima of $\chi'$ and $\chi''$. 


\begin{figure*}[t!]
\includegraphics[width= 1\textwidth]{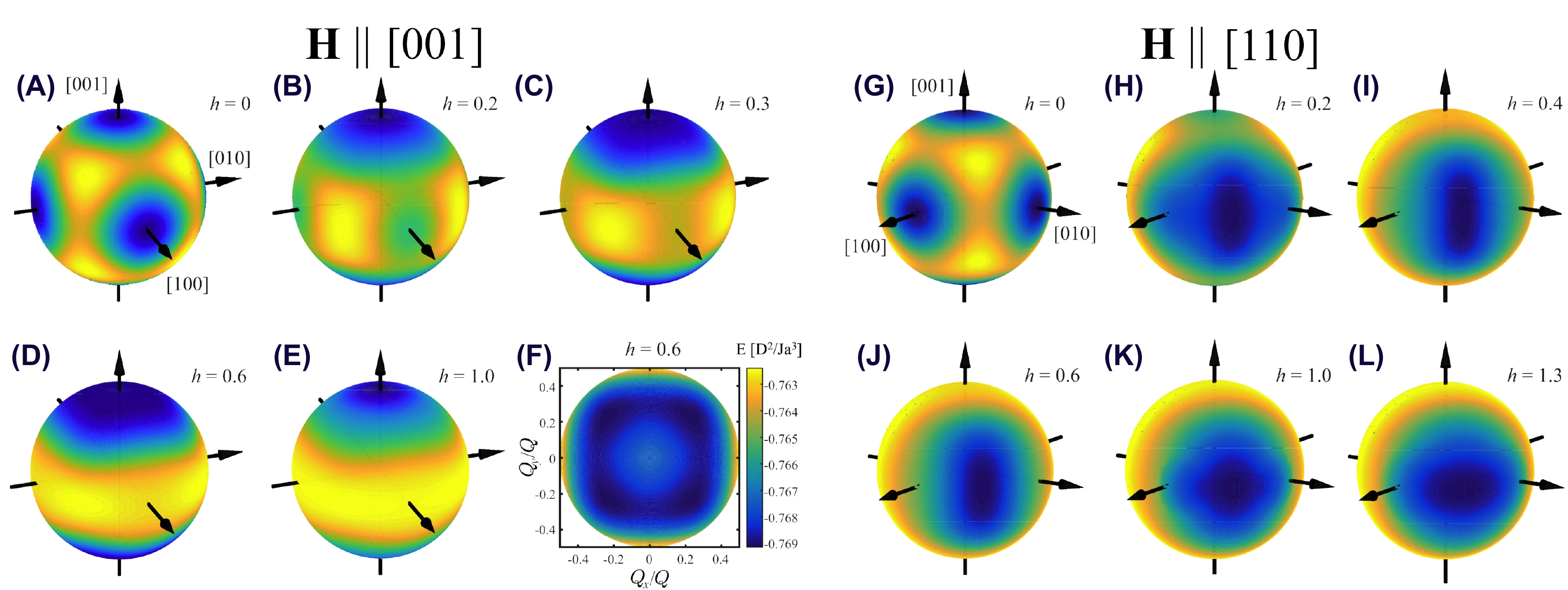}
\caption{\textbf{Q-dependence of spiral energy for $\bm H \| [001]$ and $\bm H \| [110]$}. False color plots of the conical spiral energy on the unit sphere of $\hat{\bm Q} = {\bm Q}/Q$ describing the direction of the spiral wave vector for various values of the dimensionless magnetic field, $h = H/H_{C2}$.  \textbf{(A-F)} Magnetic field applied along the $[001]$ direction: \textbf{(A)} $h = 0$, \textbf{(B)} $h = 0.2$, \textbf{(C)} $h = 0.3$ \textbf{(D)} $h = 0.6$ and \textbf{(E)} $h = 1.0$. \textbf{(F)} The energy sphere for $h = 0.6$, seen from above, showing four energy minima, which correspond to $\bm Q$ tilted away from the magnetic field vector along the $[\pm 1, \pm1,0]$ directions. \textbf{ (G-L)} Magnetic field applied along the $[110]$ direction: \textbf{(G)} $h = 0$, \textbf{(H)} $h = 0.2$, \textbf{(I)} $h = 0.4$ \textbf{(J)} $h = 0.6$, \textbf{(K)} $h = 1.0$ and \textbf{(L)} $h = 1.3$.
}  
\label{Fig:sphere}
\end{figure*}

\section*{Discussion}

%
This remarkable behavior can be understood in terms of competing  magnetic anisotropies that are generic to cubic chiral magnets. 
This competition is particularly tight in \cuse, as explained below. Despite the long history of studies of cubic chiral magnets \cite{1Nakanishi:1980SS, Bak:1980ud, Belitz:2006vd}, the discussion of anisotropic magnetic  interactions in these materials remains to our best knowledge incomplete.
 
The starting point of our approach is a continuum model with the free energy per unit cell:
\begin{equation}
{\cal E} = 
\frac{Ja^2}{2} \sum_{i=x,y,z}\partial_i \bm m \cdot \partial_i \bm m + 
Da\;  \bm m \cdot \bm \nabla \times \bm m - a^3  \mu_0 M \bm m \cdot {\bm H} + {\cal E}_{a},
\label{eq:model1}
\end{equation} 
where $\bm m$ is the unit vector in the direction of the magnetization, $M$ is the magnetization value, 
$a$ is the lattice constant, and ${\cal E}_{a}$ is the magnetic anisotropy energy. 
There  are five terms of fourth order in the spin-orbit coupling, $\lambda$, allowed by the P2$_1$3 symmetry: 
\begin{equation}
 \begin{array}{rcc} 
 {\cal E}_{\rm a} &=& 
 C_1 a^2 \left[\partial_x m_x \partial_x m_x + \partial_y m_y \partial_y m_y + \partial_z m_z \partial_z m_z\right]\\ 
 && 
 +C_2 a^2 \left[\partial_z m_x \partial_z m_x + \partial_x m_y \partial_x m_y + \partial_y m_z \partial_y m_z\right.\\
 &&
 \;\;\;\;\;\;\;-\left.(\partial_y m_x \partial_y m_x + \partial_z m_y \partial_z m_y + \partial_x m_z \partial_x m_z)\right]\\
 &&
 +2C_3 a^2 \left[\partial_x m_x \partial_y m_y + \partial_y m_y \partial_z m_z+ \partial_z m_z \partial_x m_x\right]\\
 &&
 +\frac{1}{2}{\cal J} a^4 \left[\partial_x^2 \bm m \cdot \partial_x^2 \bm m + \partial_y^2 \bm m \cdot \partial_y^2 \bm m  + \partial_z^2 \bm m \cdot \partial_z^2 \bm m\right]\\
&&
+K \left( m_x^4 + m_y^4 + m_z^4 \right). 
 \end{array}
 \label{eq:anisotropy}
 \end{equation}
Their  importance can be understood by substituting into Eq.(\ref{eq:model1}) the conical spiral Ansatz:
\begin{equation}
\bm m = \cos\theta \bm e_3 + \sin \theta \left[ \cos (\bm Q \cdot \bm x) \bm e_1 + \sin (\bm Q \cdot \bm x) \bm e_2\right],
\label{eq:Ansatz}
\end{equation}
where $\theta$ is the conical angle and  $(\bm e_1,\bm e_2,\bm e_3)$ are three mutually orthogonal unit vectors. If the magnetic anisotropy and Zeeman energies are neglected,  $Qa =\frac{D}{J}$ is independent of the orientation of ${\bm Q}$. The applied magnetic field favors $\bm Q\| \bm H$ with the conical angle given by $\cos \theta = \frac{H}{H_{C2}}$, where $\mu_0 M H_{C2} = \frac{D^2}{J a^3}$ defines the critical field, $H_{C2}$. 

The DM interaction originates from the antisymmetric anisotropic exchange between Cu spins, which is the first-order correction to the Heisenberg exchange in powers of $\lambda$: $D \sim J \zeta$, where $\zeta = \frac{\lambda}{\Delta}$, $\Delta$ being the typical electron excitation energy on Cu sites \cite{Moriya:1960go, Coffey:1992ff}. The first three anisotropy terms in Eq.(\ref{eq:anisotropy}) result from the symmetric anisotropic exchange between Cu ions and are proportional to the second power of the spin-orbit coupling\cite{Moriya:1960go, Coffey:1992ff}: $C_i \sim J \zeta^2$. Since $Qa = \frac{D}{J} = \zeta$, these anisotropy terms are of order of $J \zeta^4 $. The fourth term in Eq.(\ref{eq:anisotropy}) results from the expansion of the Heisenberg exchange interaction in powers of $Qa$ and is also $\sim J \zeta^4$. The last term in Eq.(\ref{eq:anisotropy}) has the form of the fourth-order single-ion anisotropy allowed by cubic symmetry. In absence of the single-ion anisotropy for Cu ions with $S = \frac{1}{2}$, this term emerges at the scale of the unit cell containing 16 Cu ions, which form a network of tetrahedra with $S = 1$ \cite{Bos08, Janson:2014uo}. This last term appears either as a second-order correction to the magnetic energy in powers of the symmetric anisotropic exchange or as a fourth-order correction in the antisymmetric exchange, both $\propto \zeta^4$. Importantly, the intermediate states
are excited states of Cu tetrahedra\cite{2014PhRvB..90n0404R} with energy $\sim J$ rather than the electronic excitations of Cu ions with  energy $\Delta$. As a result, also the last term is $\sim J \zeta^4$. Therefore, the magnetic block structure of \cuse\ makes all anisotropy terms in Eq.(\ref{eq:anisotropy}) comparable, which can frustrate the direction of  $\bm Q$.

%
Another important point is that the direction of  $\bm Q$, favored by a magnetic anisotropy term, may vary with the strength of the applied magnetic field, because it depends on the conical angle,  $\theta = \theta(H)$.
%
%
%
Supplementary Fig.~\ref{fig:keff} shows the $\theta$-dependence of the fourth-order effective anisotropy, $K_{\rm eff} = K B(\theta)$, which is negative for small $\theta$ and for $\theta \sim \frac{\pi}{2}$, stabilizing  the helical spiral with $\bm Q \| \langle 001 \rangle$, as it is the case for \cuse\ at zero field. 
However, for intermediate values of $\theta$, $K_{\rm eff}$ is positive and the preferred direction of $\bm Q$ becomes $\langle 111 \rangle$. 
In this interval of $\theta$, spins in the conical spiral with $\bm Q \| \langle 001 \rangle $ are closer to the body diagonals than to cubic axes, which makes this wave vector direction unfavorable (see Supplementary Sec.~\ref{sec:anisotropies} for more details).

This effect gives rise to local minima of the conical spiral energy in  $Q$-space. If only the fourth-order anisotropy is taken into account, the global energy minimum for $\bm H \| \langle 001 \rangle$ is still at $\bm Q \| \bm H$. However, additional anisotropy  terms can turn these local minima into the global ones, so that in some magnetic field interval the tilted conical spiral becomes the ground state. Fig. \ref{Fig:sphere}A-F shows the false color plot of the conical spiral energy as a function of  {\bf Q} for several values of the dimensionless magnetic field, $h = H/H_{C2}$, applied along the $[ 001 ]$ direction. For simplicity, only two anisotropy terms are nonzero in this calculation: 
$\kappa = \frac{KJ}{D^2}  = - 0.19$ and $\gamma_1 = \frac{C_1}{J} = - 0.1$. 
In zero field (Fig. \ref{Fig:sphere}A), there are three energy minima along $\langle 001 \rangle$ i.e. along the [001], [010] and [001] directions, corresponding to three degenerate helical spiral domains in \cuse. For $h = 0.2$ (Fig. \ref{Fig:sphere}B), the helical spiral states with $\bm Q$ along the  [001]  and [010] directions are metastable. For $h = 0.3$ (Fig. \ref{Fig:sphere}C) only the minimum with $\bm Q \| [001]$ exists, corresponding to the conical spiral state. For $h = 0.6$ ((Fig. \ref{Fig:sphere}D) the conical spiral with $\bm Q \| [001]$ is unstable and there appear four new minima, corresponding to four domains with $\bm Q$ tilted away from the magnetic field vector along the $[\pm 1, \pm1,0]$ directions, as can be seen more clearly in Fig. \ref{Fig:sphere}F showing the energy sphere seen from above. The relative energy changes in this part are $\sim 10^{-2}$, implying large fluctuations of the spiral wave vector, which can explain the diffuse scattering shown in Fig. \ref{Fig:SANS}D. Finally, Fig. \ref{Fig:sphere}E shows the field-polarized state at $h = 1$. 

Fig.~\ref{Fig:sphere}G-L shows the $\bm Q$-dependence of the conical spiral energy in several magnetic fields along the  $[ 110 ] $ direction, calculated for the same values of parameters as Fig.~\ref{Fig:sphere}A-F. As the magnetic field increases, the helical spiral states with $\bm Q$ along the $[ 001 ] $ and $[010]$ directions merge into a single state with the wave vector parallel to $\bm H$ and the state with $\bm Q \| [001]$ ultimately disappears (see Fig.~\ref{Fig:sphere}G,H,I). This gives rise to the two step transition from the helical to the conical phase observed experimentally. For this field direction the multi-domain tilted spiral state does not appear and there is only one global minimum with $\bm Q \| \bm H \| [ 110 ] $ for $H > H_{C1}$. Nevertheless, one can see the strong vertical elongation of the energy contours in Fig.~\ref{Fig:sphere} H,I and J, which is a result of the competition with the $\bm Q \| [11\pm1]$ states. At $h = 1$ the elongation changes from vertical to horizontal (see Fig.~\ref{Fig:sphere}K,L).
 
\begin{figure}
\includegraphics[width=\columnwidth]{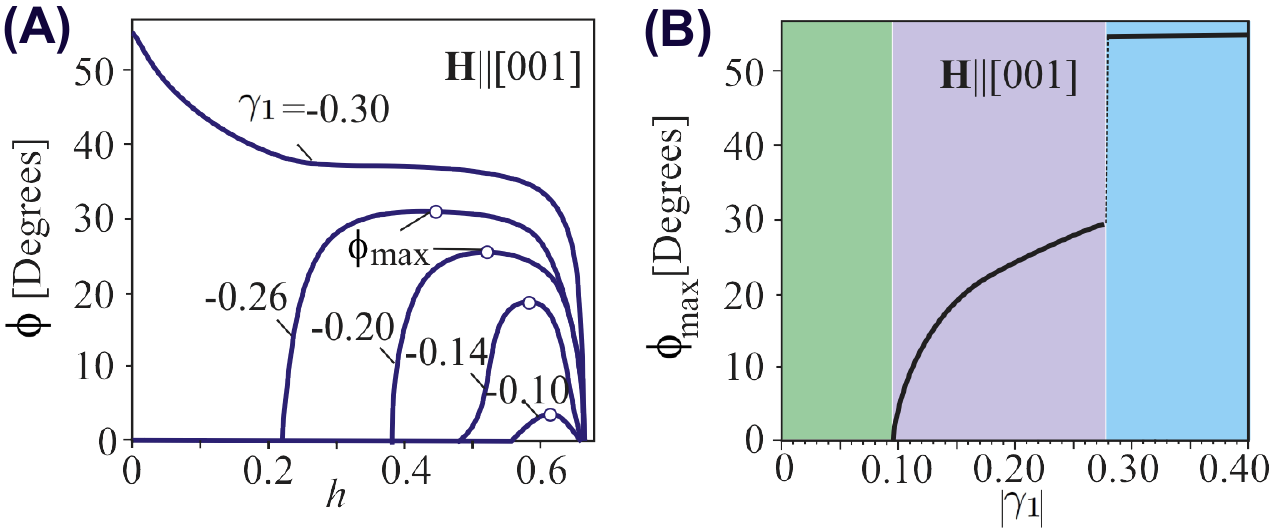}
\caption{\textbf{Field-induced re-orientation of the spiral wave vectors for the fourth-order anisotropy $\bm{\kappa}$ = -0.2 and the anisotropic exchange $\mathbf{\gamma_1<0}$.} 
{\bf (A)} Magnetic field dependence of the angle $\phi$ describing the tilt of the spiral wave vector $\bm Q$ away from $\bm{H}\|[001]$ towards the body diagonals for different values of  $|\gamma_1|$.
%
Above the critical value, $|\gamma_1| \approx 0.1$, $\phi$ increases from 0, that corresponds to the conical spiral phase,  to a maximal value, $\phi_{max}$ (empty circle), and then decreases back to 0 as the magnetic field strength, $h$, increases. For  $|\gamma_1| \gtrsim 0.28$, $\bm{Q}\| \langle 111 \rangle$ even at  zero field   and smoothly rotates towards [001] with  increasing field. 
%
%
{\bf (B)} Dependence of $\phi_{max}$.
For $|\gamma_1|\lesssim 0.1$ (green area), $\phi_{max}$=0, which implies that  $\bm{Q}\|\langle100\rangle$. By increasing the strength of  the anisotropic exchange  $\bm{Q}$ smoothly rotates towards $\langle111\rangle$ (purple area) and for $|\gamma_1| \gtrsim 0.28$, $\bm{Q}\|\langle111\rangle$ (blue area). 
\label{angles}
}
\end{figure}

 %
 These conclusions drawn using the variational approach are confirmed by exact energy minimization of  Eq.(\ref{eq:model1}) including two competing anisotropy terms: the fourth-order anisotropy with $\kappa =-0.2$ and the anisotropic exchange $\gamma_1$.
%
The case of  $\bm{H}||[110]$ is treated in the Supplementary Sec.~\ref{sec:numerics}, which explains the two  lines,  $\mu_0 H_{C1}^{(1)}$ and $\mu_0 H_{C1}^{(2)}$, for the transition from the helical to the conical spiral state, shown in Fig.~\ref{Fig:BT}A-C. 
For $\bm{H}\|[001]$, 
%
the field dependence of the angle, $\phi$,  between  $\bm Q$ and the [001] cubic axis, which describes the tilt of $\bm Q$  towards  the $\langle111\rangle$ directions is shown in Fig.~\ref{angles}A. The  tilted spiral state appears when  $|\gamma_1|$ exceeds a critical value, which is slightly lower than 0.1 for $\kappa = -0.2$. 
When the magnetic field increases, the tilt angle reaches its maximal value,  $\phi_{\rm max}$, marked by the empty circles in Fig.~\ref{angles}A and then decreases to 0.  
%
%
%
As shown in   Fig.~\ref{angles}B,  $\phi_{\rm max}$ = 0   for $|\gamma_1| \lesssim 0.1$. Thus, the state  with $\bm{Q}\| \langle100\rangle$ is stable at low anisotropies. However, as the  exchange anisotropy increases an intermediate state occurs, and finally for $|\gamma_1| \gtrsim 0.28$,  the state with $\bm{Q}\| \langle111\rangle$ is stabilized even  at zero magnetic field.  

%
%

In our diffraction experiment we do not observe all four tilted spiral domains, which is likely related to a small misalignment of the sample: due to a weak dependence of the spiral energy on $\bm Q$, even a tiny deviation of $\bm H$ from the [001] direction leads to the selection of one of the four domains. This suggests that the domain structure of the tilted state can also be controlled by an applied electric field using the multiferroic nature of Cu2OSeO3 \cite{Seki12, mochizuki2015dynamical, ruff2015magnetoelectric}. The electric polarization induced by the tilted spiral with the spin rotation axis  $\bm{l} = (\frac{1}{\sqrt{2}}\sin \alpha, \frac{1}{\sqrt{2}}\sin \alpha, \cos \alpha)$ is given by:
\begin{equation}
\langle \bm{P} \rangle = \lambda \frac{(3 \cos^2 \theta -1)}{4\sqrt{2}}\left(\sin 2\alpha, \sin 2\alpha, \frac{1}{\sqrt{2}}(1 - \cos 2\alpha)\right).
\label{eq:Pav}
\end{equation} 
For small anisotropies, $\alpha$ is close to the tilt angle of  $\bm Q$, $\phi$ (more precise relation between $\alpha$ and $\phi$ and the derivation of Eq.(\ref{eq:Pav}) can be found in Supplementary Sec.~\ref{sec:polarisation}). Since $\phi$ does not exceed 30$^\circ$, the induced electric polarization is almost normal to the applied magnetic field $\bm{H}\|[001]$.  The conical spiral with $\alpha = 0$ induces no electric polarization.

To summarize, we observe a new high-field multi-domain magnetic state which intervenes between the conical spiral and field-polarized phases and is stable in a broad temperature range. Its modulation vector can have four directions tilted away from the magnetic field vector applied along the high-symmetry [001] direction. We show that this state is a result of the interplay between different magnetic anisotropies, generic to cubic chiral magnets. Competing anisotropies can have more general implications, such as the orientation of skyrmion tubes and the transition from the triangular to square skyrmion lattice\cite{Karube2016, Nakajima:2017jx}. They can also play a role in the emergence of the partial order in MnSi under pressure\cite{Pfleiderer:2004cm}. The transition into the tilted conical spiral phase should have many observable consequences, such as the spin-Hall magnetoresistance\cite{Aqeel2016},  modified spin-wave spectrum and a magnetically-induced electric polarization.


\bibliographystyle{Science}

\bibliography{Ref_CuSe_suscep_lowT}

\section*{Acknowledgments}
The authors acknowledge fruitful discussions with Alex Bogdanov.  Dr. G. Blake and Dr. Y. Prots are acknowledged for orienting  the single crystals. FQ acknowledges financial support from the China Scholarship Council.  CP and LJB acknowledge  NWO Groot Grant No. LARMOR 721.012.102. CP and MM acknowledge  Vrije FOM-programma "Skyrmionics".  AOL  thanks Ulrike Nitzsche for technical assistance and acknowledges  JSPS Core-to-Core Program, Advanced Research Networks (Japan) and JSPS Grant-in-Aid for Research Activity Start-up 17H06889.


\section*{Supplementary materials}
\section{Materials and Methods}

Magnetization and magnetic susceptibility measurements were performed on two single crystals of  \cuse~with dimensions of $\sim{1}\times{1}\times{1}$~mm$^{3}$ grown at the Zernike Institute for Advanced Materials. One crystal was  oriented with a $\langle001\rangle$ axis vertical while the other one was oriented with a $\langle 110 \rangle $ axis vertical. 
A third single crystal with dimensions $\sim{3} \times{3} \times{4} $~mm$^{3}$ grown at the Max Planck Institute for Chemical Physics of Solids was used for the neutron scattering measurements. This sample was oriented with the $[ \bar{1}10 ] $ crystallographic axis vertical. 
All crystals were prepared by  chemical vapor transport method and their quality and structure was checked by x-ray diffraction. 

Magnetization and magnetic susceptibility  were measured with a MPMS-5XL SQUID using the extraction method. For the determination of the magnetization a static magnetic field, $\mu_0 H$, was applied along the vertical direction. The real  and imaginary parts of the magnetic ac susceptibility,  $\chi'$  and $\chi''$,  were measured by adding to $\mu_0 H$ a vertical drive ac field, $\mu_0 H_{ac}$, with an amplitude of $0.4$~mT. The frequency of  $H_{ac}$ was varied between $0.1$ and $1000$~Hz.

The SANS measurements were performed on the instruments PA$20$ of the Laboratoire L\'{e}on Brillouin and GP-SANS of Oak Ridge National Laboratory using  neutron wavelengths of 0.6~nm and 1~nm respectively. At both instruments the magnetic field was applied either parallel or perpendicular to the incoming neutron beam wave vector $\bm k_i$ using a horizontal magnetic field cryomagnet. 
The orientation of the crystal axes  with respect to $\bm k_i$ and to the magnetic field was varied by rotating the sample  in the cryomagnet.
The SANS patterns were collected for $\bm H \| [ 110 ] $ and $\bm H \| [ 001]$ and, in each case, for $ \bm H \bot \bm k_i$ and $\bm H \| \bm k_i$, by rotating both the sample and the magnetic field through $90^\circ$ with respect to $\bm k_i$.  
Measurements at $70$~K, where the magnetic scattering is negligible, were used  for the background correction of the SANS patterns.

All measurements have been performed after zero field cooling the sample through the magnetic transition temperature, $T_c$, down to the temperature of interest. 
The magnetic field was then increased stepwise.
The applied magnetic field, $\mu_0 H_{ext}$ , was corrected for the demagnetizing effect to obtain the internal magnetic field, $\mu_0H_{int}$ (in SI units):
\begin{equation}
\bm H_{int}= \bm H_{ext}-N {\bm M},
\end{equation}
\noindent where $N = 1/3$ is the demagnetization factor for our  (nearly) cubic shape samples
The demagnetizing field correction also modifies the values of the  magnetic susceptibility:
\begin{equation}
\chi_{int}=  \frac{\chi_{ext}}{1-\mu_0 N \chi_{ext}}
\end{equation}


\section{Supplementary Text}

\begin{figure*}[t]
\includegraphics[width= 0.9\textwidth]{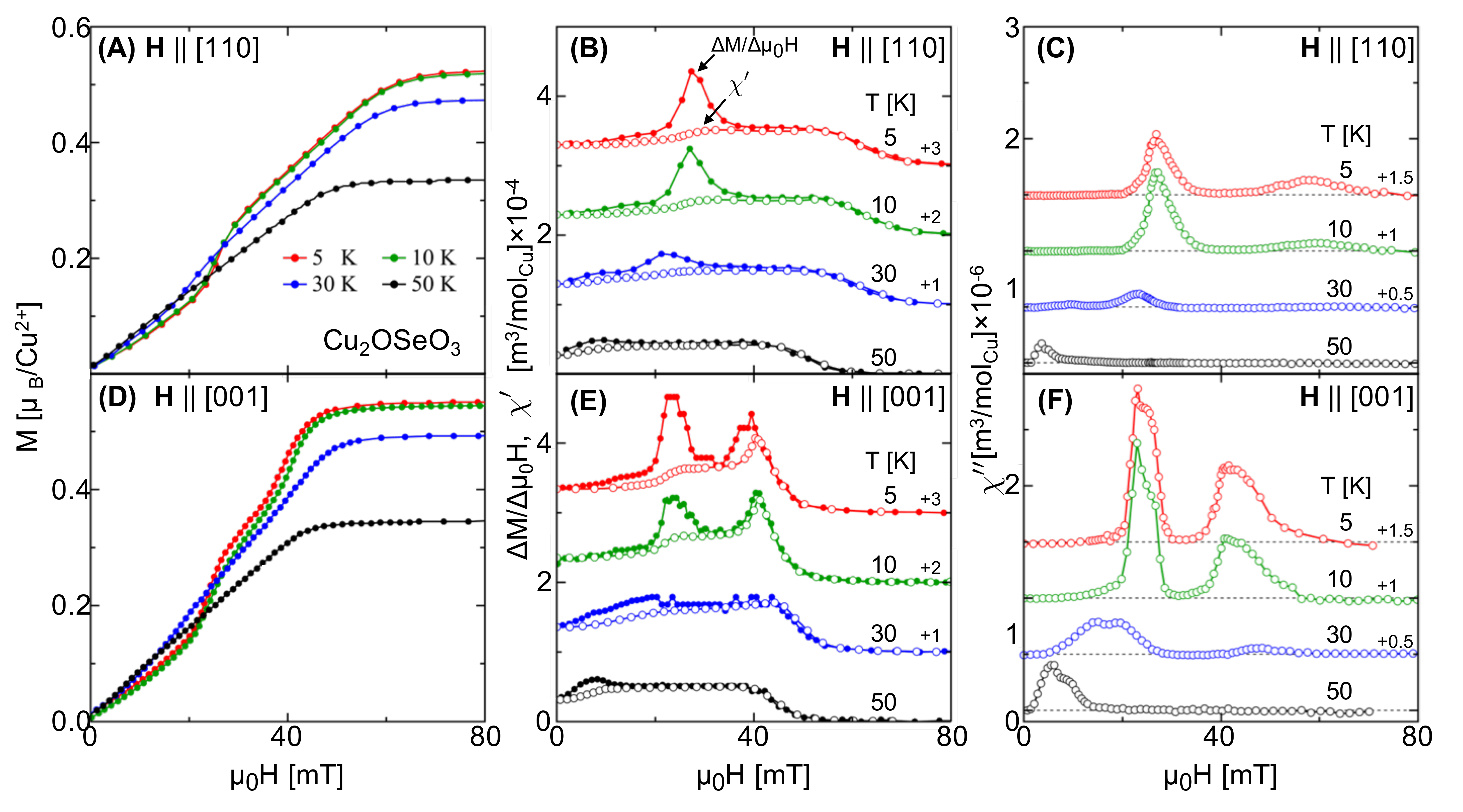}
\caption{\textbf{Magnetic properties of \cuse}. Magnetic field and temperature dependence of the magnetization $M$, its derivative, $\Delta M/\Delta (\mu_0 H)$, and the real and imaginary parts of the ac magnetic susceptibility, $\chi'$ and $\chi''$,  at a frequency of 10~Hz for $\bm H \| [ 110 ] $ ({\bf A-C}) and $\bm H\| [ 001 ]$ ({\bf D-F}).  The $\Delta M/\Delta (\mu_0 H)$, $\chi'$ and $\chi''$ curves are shifted vertically with respect to the base line  by the values indicated.} 
\label{Fig:ac}
\end{figure*}
\begin{figure}[b]
\includegraphics[width= 0.9\columnwidth]{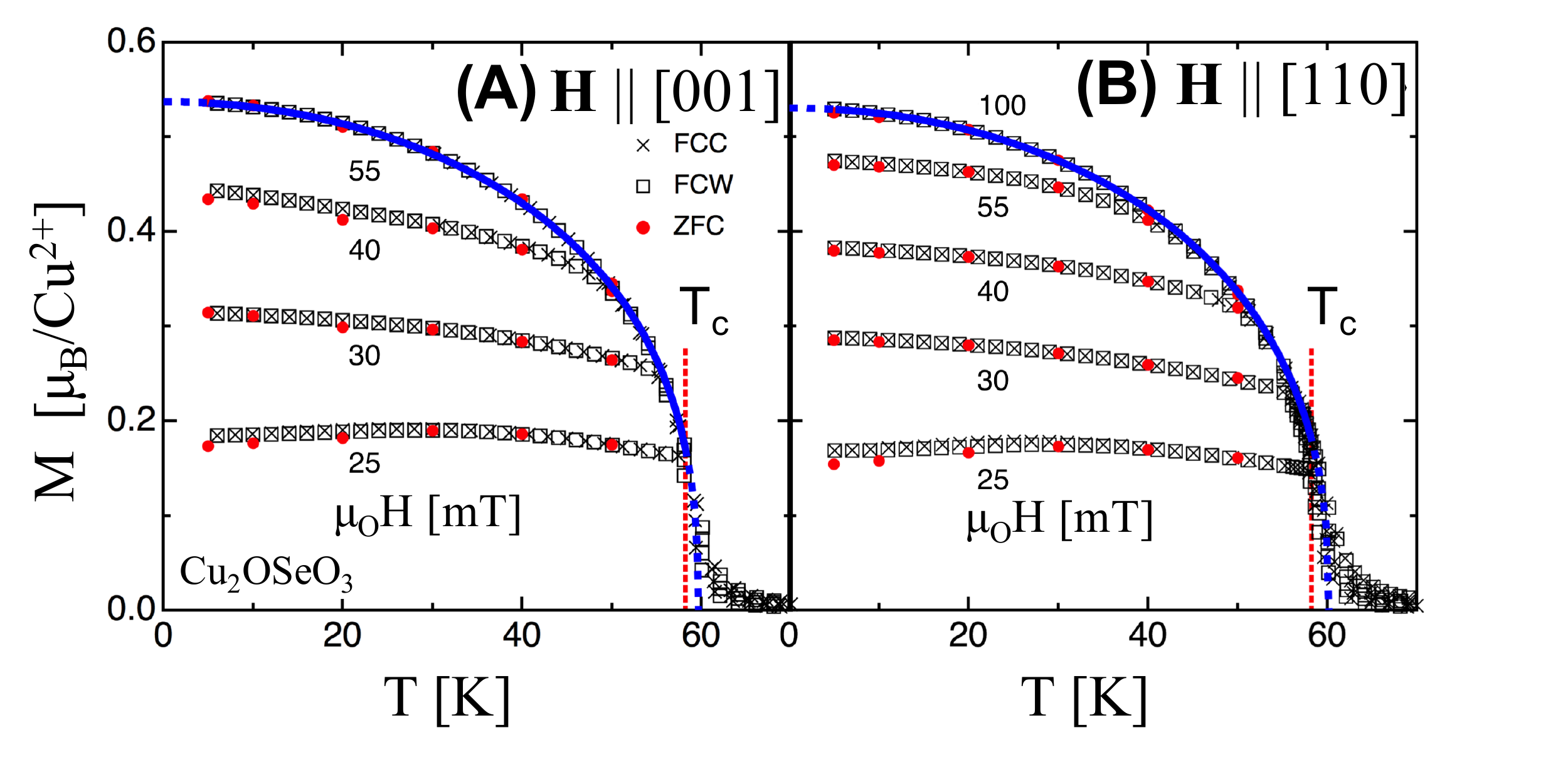}
\caption{  \textbf{Temperature dependence of the magnetization}. The zero field cooled (ZFC) and field cooled magnetization was measured by  either cooling (FCC) or warming (FCW) the sample, for $\bm H \| [{001}]$ ({\bf A}) and  $\bm H \|[{110}]$ ({\bf B}).  The blue lines in both panels are fits of eq.~\ref{eq:msat} with the parameters given in the Table \ref{Tbl:powerlaw}. The red vertical dashed lines indicate $T_c = {58.2}$~K} 
\label{Fig:MvsT.pdf}
\end{figure}

\begin{figure}[b]
\includegraphics[width= 0.9\columnwidth]{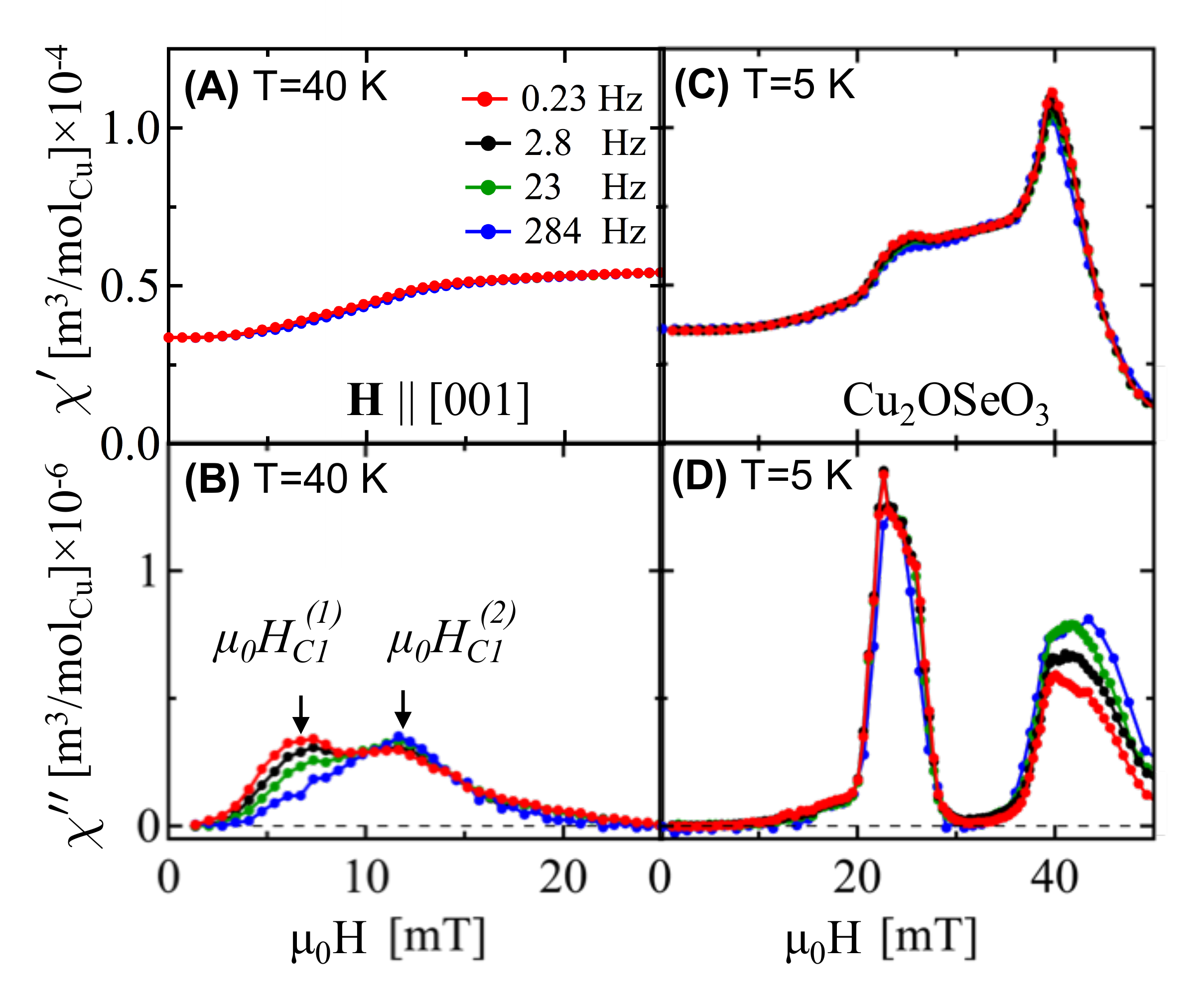}
\caption{  \textbf{$\chi'$ and $\chi''$ as a function of magnetic field}. The results are given for various frequencies at T = $40$~K in panels ({\bf A}) and ({\bf B}), and T = $5$~K in panels ({\bf C}) and ({\bf D}). The magnetic field was applied along $[ 001 ]$.} 
\label{Fig:varfvsB}
\end{figure}

\subsection{Temperature and Magnetic Field Dependence of the Magnetization and Susceptibility}
\label{sec:susc}

Fig. \ref{Fig:ac} shows the low temperature magnetic field and temperature dependence of  the magnetization $M$, its derivative, $\Delta M/\Delta (\mu_0 H)$, and the real and imaginary parts of the ac magnetic susceptibility, $\chi'$ and $\chi''$. Whereas for  $\bm H \| [ 110 ]$  $M$ increases almost linearly with magnetic field until saturation, clear deviations from this linear increase occur for  $\bm H \| [ 001 ]$. These are highlighted by the  emergence of peaks in  the derivative, $\Delta M/\Delta (\mu_0 H)$, $\chi'$ and $\chi''$, which are most prominent at low temperatures and indicate  complex rearrangements of spins discussed below.
At low magnetic fields, where the transition between the helical and conical spiral states is expected, $\Delta M/\Delta (\mu_0 H)$ shows a maximum, which is not seen in  $\chi'$. This discrepancy as well as the  corresponding strong $\chi''$ signal indicate strong frequency dependence and  energy dissipation. 
A closer inspection of the frequency dependence presented below reveals that this  $\chi''$ signal is composed of two adjacent peaks, each with its own frequency dependence.
At higher magnetic fields, just below the saturation of  magnetization, additional peaks appear in $\Delta M/\Delta (\mu_0 H)$, $\chi'$ and $\chi''$ but only for $\bm H \| [ 001 ]$ and  $T < 30$~K. 
These peaks are centered at $\sim$45~mT  and  $\sim$40~mT  at 30~K and 5~K, respectively, and are  associated with the new unexpected state in the phase diagram of Cu$_2$OSeO$_3$  revealed by  SANS.

Fig.~\ref{Fig:MvsT.pdf} displays the temperature dependence of the  magnetization for some specific magnetic fields with $\bm H \| [{001}]$ in A and  $\bm H \| [{110}]$ in B. Besides the  ZFC data from the $M$ vs $\mu_0 H$ curves the figure also includes magnetization points obtained by cooling the sample under magnetic field through $T_c$, thus following a ``field cooled'', FC, protocol. These magnetization points have been measured  by either cooling (FCC) or warming the sample (FCW) and do not show any noticeable dependence on the magnetic field history. The data  follow a generic envelop, which is accounted for by the modified power law \cite{Zivkovi12}: 
\begin{equation}
\label{eq:msat}
 M_{sat}(T) \; = \; M_{sat}(0) \; \left[1-\left(T/T_c^o\right)^n \right]^\beta ,
\end{equation}
\noindent with $M_{sat}({0})$ the saturated magnetic moment at $T={0}$~K, $n$ and $\beta$  critical exponents, and $T_c^o$ the associated critical temperature. The fit leads to the continuous lines in Fig.~\ref{Fig:MvsT.pdf}. The fitting parameters  are tabulated in Table~\ref{Tbl:powerlaw} and are almost identical for the two sample orientations  and  in agreement with a previous study  \cite{Zivkovi12}. Close to $T_C$, Eq.~\ref{eq:msat} reduces to the simple power law with $n= {1}$), that is predicted theoretically \cite{Janson:2014uo} and mimics a $3$D-Heisenberg behavior. Deviations from these generic curves occur at the $\mu_0 H_{C2}(T)$ line. 

\begin{table*}[t]
\caption{ \textbf{Temperature variation of $\bm{M_{sat}}$}.  Parameters derived from the fit of Eq. \ref{eq:msat} to the temperature dependence of $M_{sat}$ below $T_C \,= \,58.2$~K.  }
	\begin{tabular} { p{2.5cm} | p{2.5cm}  p{2.5cm} p{2.5cm} p{2.cm} }
	\hline
	\hline
	&~$M_{sat}(0)$ &$T_c^o-T_c$&$\beta$&~$n$ \\
	\hline
	 ~~~$\bm H \parallel[{001}]$& ~0.54 (1) &1.5 (1)&0.368 (5) &1.99 (2)  \\
	~~~$\bm H \parallel[{110}]$ & ~0.53 (1) & 1.8 (3)  & 0.391 (2)  & 2.01 (1)  \\
	~~~Ref[\citenum{Zivkovi12}] & ~0.559 (7) & 1.8 (1)  & 0.393 (4)  &1.95 \\
	\hline
	\hline
	\end{tabular}
\label{Tbl:powerlaw}
\end{table*}

 \subsection{Frequency dependence of the ac susceptibility}
The frequency dependence of the ac susceptibility is most visible for $\bm H \| [ 001 ]$ and in the following we will focus on the results obtained in this orientation. Fig.~\ref{Fig:varfvsB}  displays $\chi'$ and $\chi''$ measured at the indicated frequencies at $40$ and $5$~K. For both temperatures, $\chi'$ varies weakly with frequency and the frequency dependence is mainly visible for $\chi''$. At $40$~K, $\chi''$ shows two peaks which are located at $7$ and $12$~mT, thus close to the $\mu_0 H_{C1}$(T) line. The amplitude of the peak at $7$~mT decreases with increasing frequency, while the one at $12$~mT slightly increases. At a frequency of $284$~Hz only the peak at $12$~mT is visible. At $5$~K these two $\chi''$ peaks  seem to merge and vary very weakly with frequency. At this temperature a noticeable frequency dependence is  found only for the broad peak of $\chi''$ at $\sim$ 40~mT, that corresponds to the tilted spiral state.

\begin{figure}
\includegraphics[width= 0.9\columnwidth]{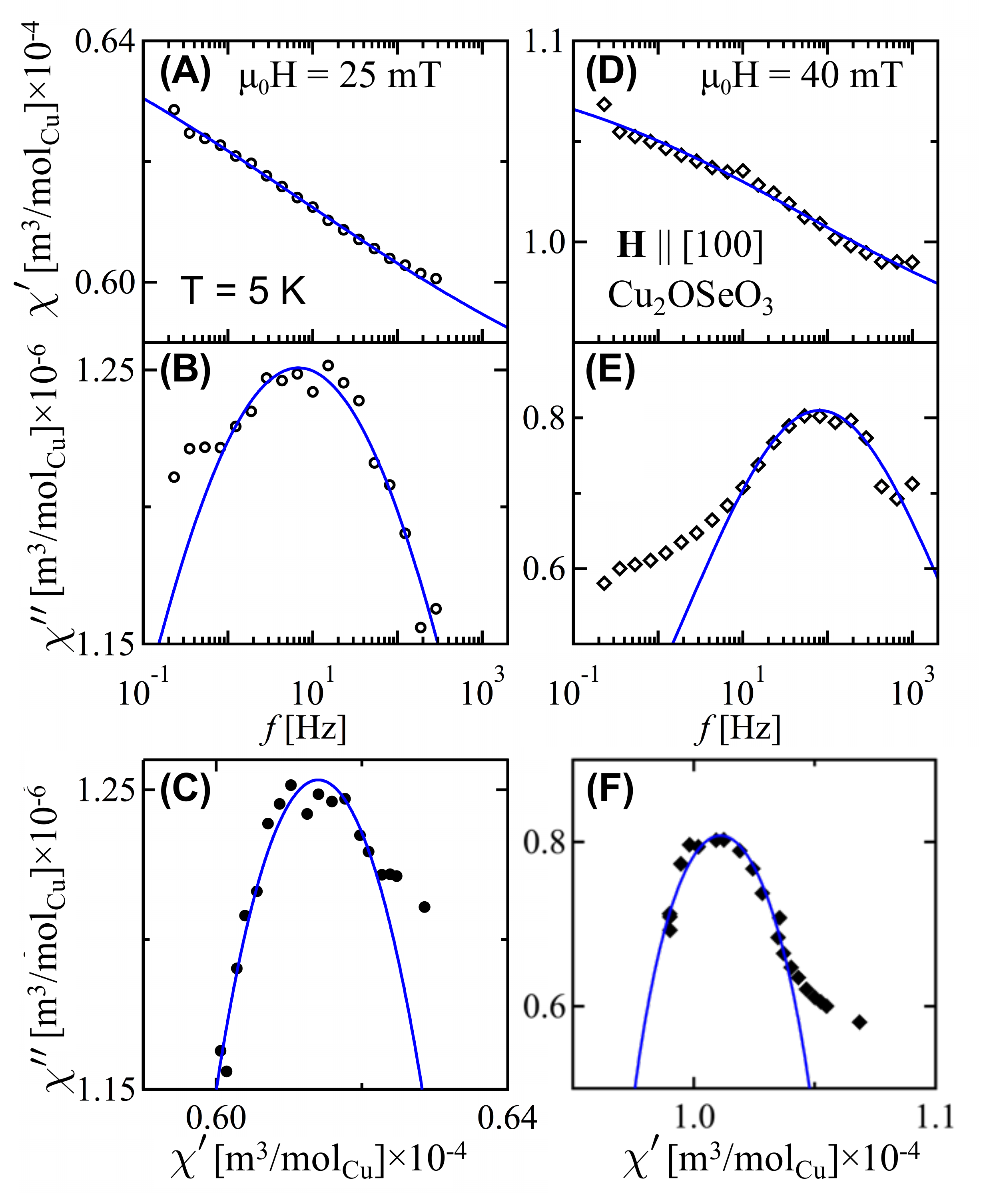}
\caption{\textbf{Frequency dependence of $\chi'$ and $\chi''$ at T = 5~K}. The data have been collected at  25~mT ({\bf A}, {\bf B}) and 40~mT ({\bf D}, {\bf E}) and for $\bm H \|  [001]$. The respective Cole-Cole plots of $\chi''$ versus $\chi'$ are given in  {\bf C}, {\bf F} where the solid lines represent fits to the  Cole-Cole formalism.} 
\label{Fig:cole}
\end{figure}

For a quantitative analysis of the frequency dependence of $\chi'$ and $\chi''$ we considered the Cole-Cole formalism modified to include a distribution of characteristic relaxation times discussed in our previous report  \cite{Qian2016}:
\begin{equation}
\chi(\omega) \; = \; \chi(\infty)+\frac{\chi(0)-\chi(\infty)}{1+(i\omega \tau_0)^{1-\alpha}},
\label{eq:colecole_start}
\end{equation}
\noindent with $\omega={2}\pi f$ the angular frequency, $\chi(0)$ and $\chi(\infty)$ the adiabatic and isothermal susceptibilities respectively and $\tau_0={1}/({2}\pi f_0)$ the characteristic relaxation time with $f_0$ the characteristic frequency. The parameter $\alpha$ measures the distribution of characteristic relaxation times with $\alpha={0}$ for a single relaxation time, and $\alpha>{0}$ for a distribution of relaxation times, which becomes broader as $\alpha$ approaches $1$. Eq.~\ref{eq:colecole_start} leads to the in-phase and out-of-phase components of the susceptibility  discussed in our previous report  \cite{Qian2016}. 

Fig.~\ref{Fig:cole} depicts the frequency dependence of ZFC $\chi'$ and $\chi''$ at $5$~K and for $\mu_0 H$ = 25 and 40~mT. For both magnetic fields, $\chi'$ varies  only slightly with frequency and decreases by no more than $\sim3\%$ over almost four orders of magnitude in frequency.  The most significant variation is found for $\chi''$, for which broad maxima develop around ${10}$~Hz and $100$~Hz for $25$ and $40$~mT respectively. 

The solid lines in Fig.~\ref{Fig:cole} illustrate a fit of the data to the Cole-Cole formalism. It leads to high values of $\alpha$, $\sim${0.85} and $\sim${0.65} for $B=25$ and $40$~mT respectively, which reflect the existence of broad distributions of relaxation times. Furthermore, deviations from the simple Cole-Cole behavior occur at low frequencies indicating  the presence of additional relaxation processes. This is consistent with the behavior  close to $T_c$  \cite{Qian2016}. 

\subsection{Phase boundaries determined from SANS}

\begin{figure}
\includegraphics[width= 0.9\columnwidth]{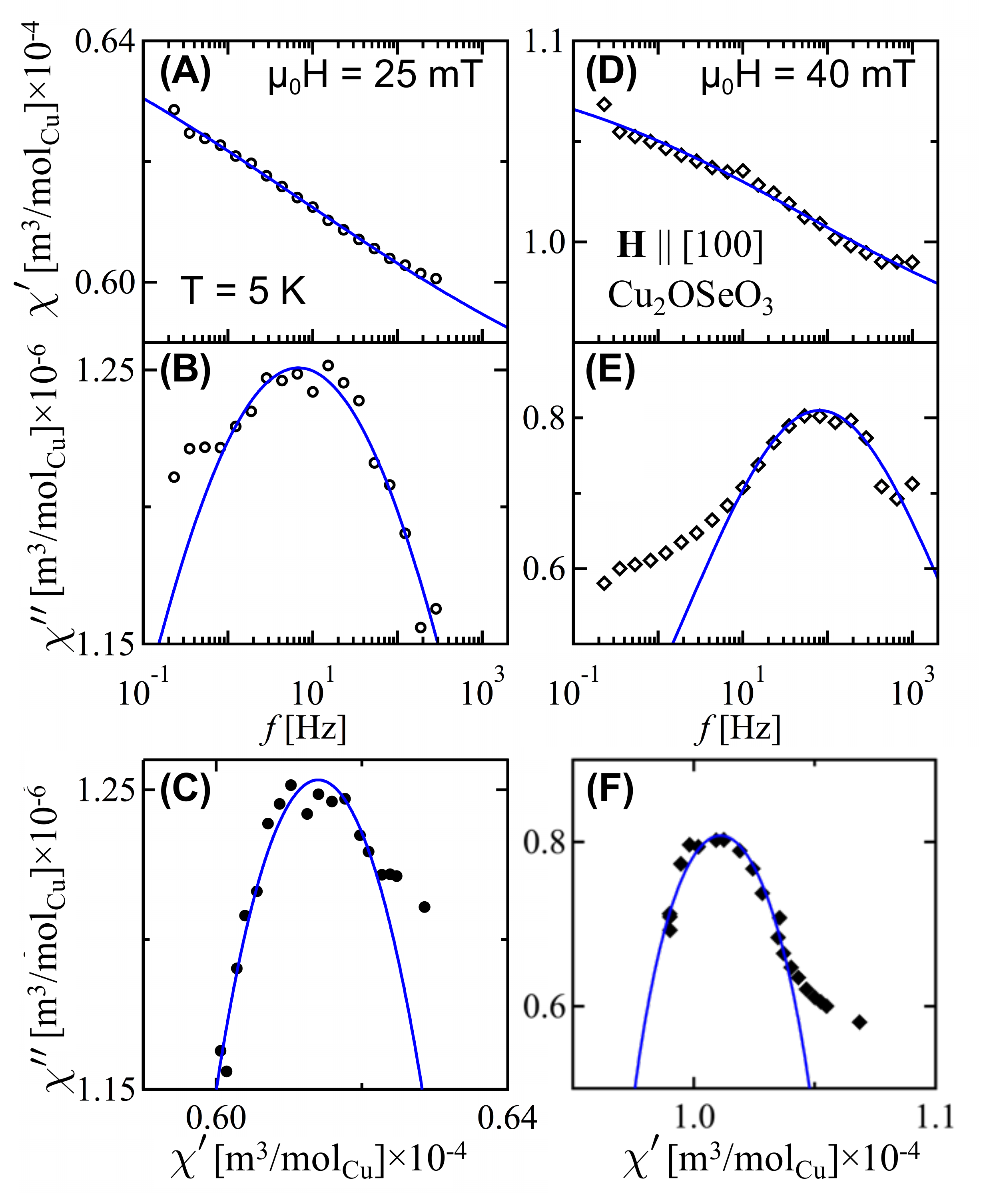}
\caption{  \textbf{Magnetic field dependence of the SANS intensity at 6~K}. The data have been obtained by integrating over the entire detector for $\bm H \| [ 001 ]$ ({\bf A}) and $\bm H \| [ 110 ]$ ({\bf B}) and for $\bm H \bot \bm k_i$ (red squares) and $\bm H \| \bm k_i$ (black circles).  The solid and open red squares correspond to data recorded by stepwise increasing and decreasing the magnetic field respectively.} 
\label{Fig:suppSANS}
\end{figure}

Fig.~\ref{Fig:suppSANS} shows the magnetic field dependence of the total scattered intensity, integrated over the detector (thus over all peaks), at $T = 6$~K. The figure shows data obtained for $\bm H \bot \bm k_i$ and collected by stepwise increasing the magnetic field. Data were also collected by stepwise decreasing the magnetic field. Both data sets are very close to each other and thus no substantial hysteresis has been found,

In the configuration $\bm H \| [ 001 ]$ and  $\bm H  \| \bm k_i$ the  intensity  decreases sharply with increasing magnetic field  and almost vanishes around  $20$~mT, which defines the $\mu_0 \bm H_{C1}^{(1)}$ transition point on the phase diagram of Fig. \ref{Fig:BT}F. The increase of intensity at higher magnetic fields originates from the diffuse scattering of the tilted spiral phase, which is most pronounced around 40~mT. In the complementary configuration of $\bm H\bot \bm k_i$   the total scattered  intensity starts to decrease above 30~mT, i.e. in the titled spiral phase, and disappears around 50~mT, which corresponds to  the $\mu_0$H$_{C2}$ transition point  on the phase diagram of Fig.~\ref{Fig:BT}. In the configuration $\bm H \| [ 110 ]$ and  $\bm H \| \bm k_i$  a similar analysis  leads to the determination of  the $\mu_0 H_{C1}^{(1)}$ transition line. However, the $\mu_0 H_{C2}$ point could not be determined as it lies at  magnetic fields that exceed the window of these measurements. 
\begin{figure}[h]
\includegraphics[width= 0.9\columnwidth]{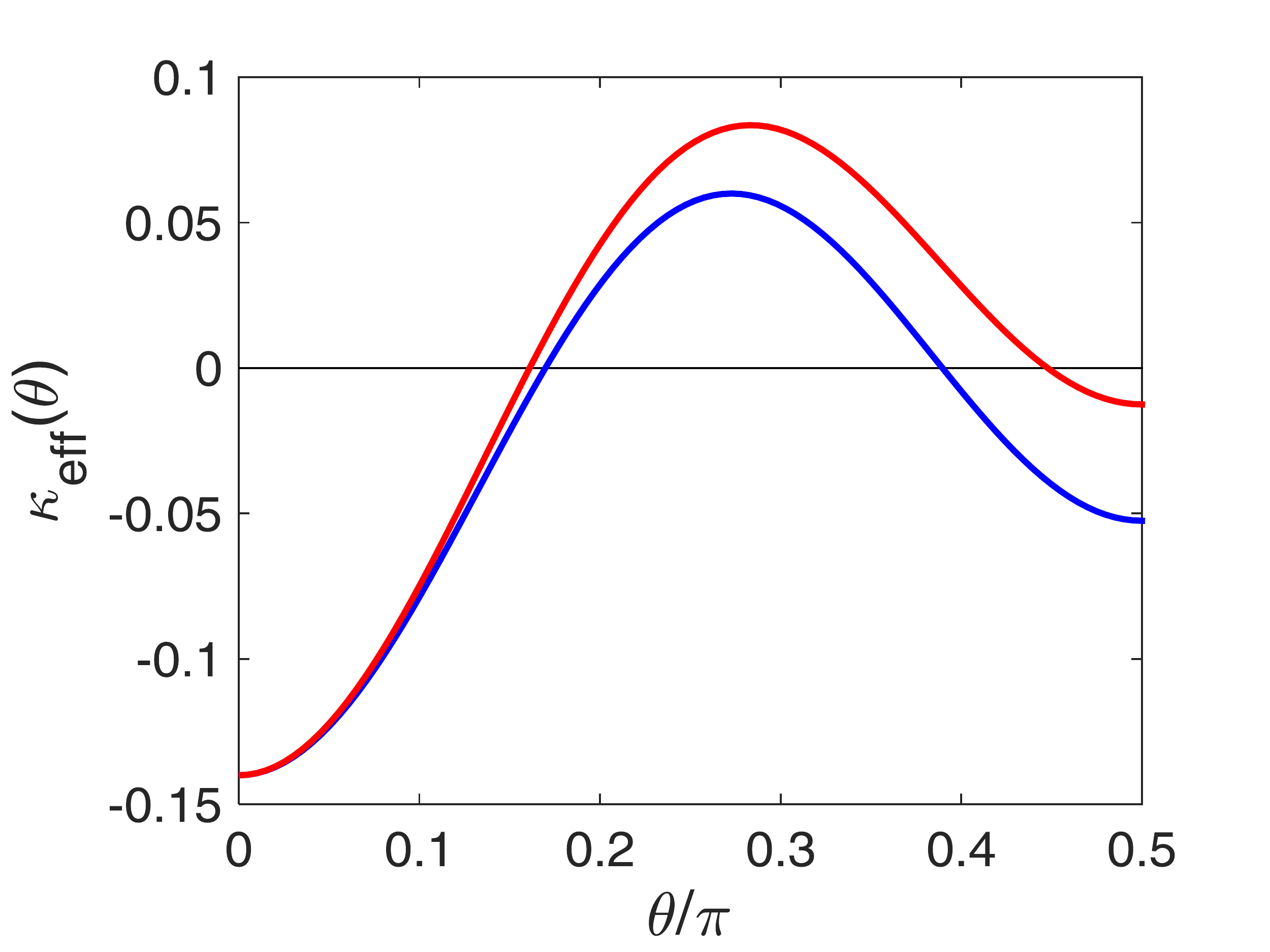}
\caption{\textbf{Effective anisotropy.} Effective dimensionless anisotropy for the spin rotation axis, $\kappa_{\rm eff} \,= \, \kappa B(\theta)$, vs the conical angle, $\theta$, for $\kappa = \frac{K J}{D^2}=-0.14$  (blue line) and all other magnetic anisotropies equal zero. Red line shows $\kappa_{\rm eff}$ when an additional anisotropy $\gamma_1 =  \frac{C_1}{J} =  -0.08$ is added. \label{fig:keff}} 
\end{figure} 
\begin{figure}[t!]
\includegraphics[width=0.9\columnwidth]{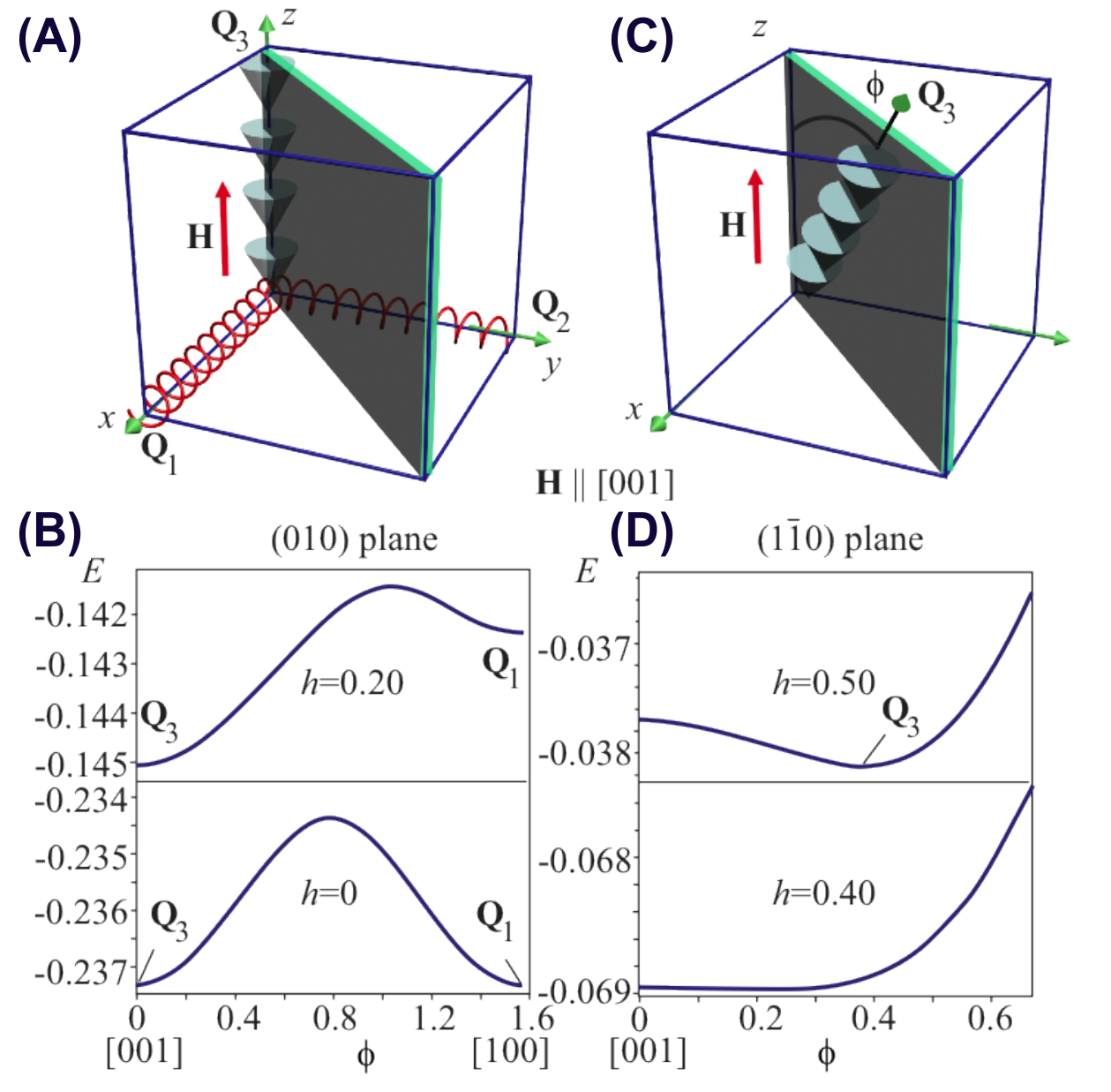}
\caption{\textbf{Spiral re-orientation for $\bm{H}\|[001]$}. ({\bf A}) Three spiral domains in low magnetic fields. 
({\bf B}) $\phi$-dependence of the spiral energy density for $\bm Q$ varying in the (010) plane in zero field and at $h = 0.2$. The global energy minimum is reached at $\bm Q_3\|[001]$ ($\phi$ is given in radians).
({\bf C}) Tilted-spiral state that appears above a critical value of the magnetic field, in which $\bm Q$ rotates away from the magnetic field vector towards the $\langle111\rangle$  directions.
({\bf D})
Energy density plotted as a function of the tilt angle in the $(1\overline{1}0)$ plane for $h = 0.4$ and $h = 0.5$. 
\label{h001}
}
\end{figure}

\begin{figure}[t!]
\includegraphics[width=0.9\columnwidth]{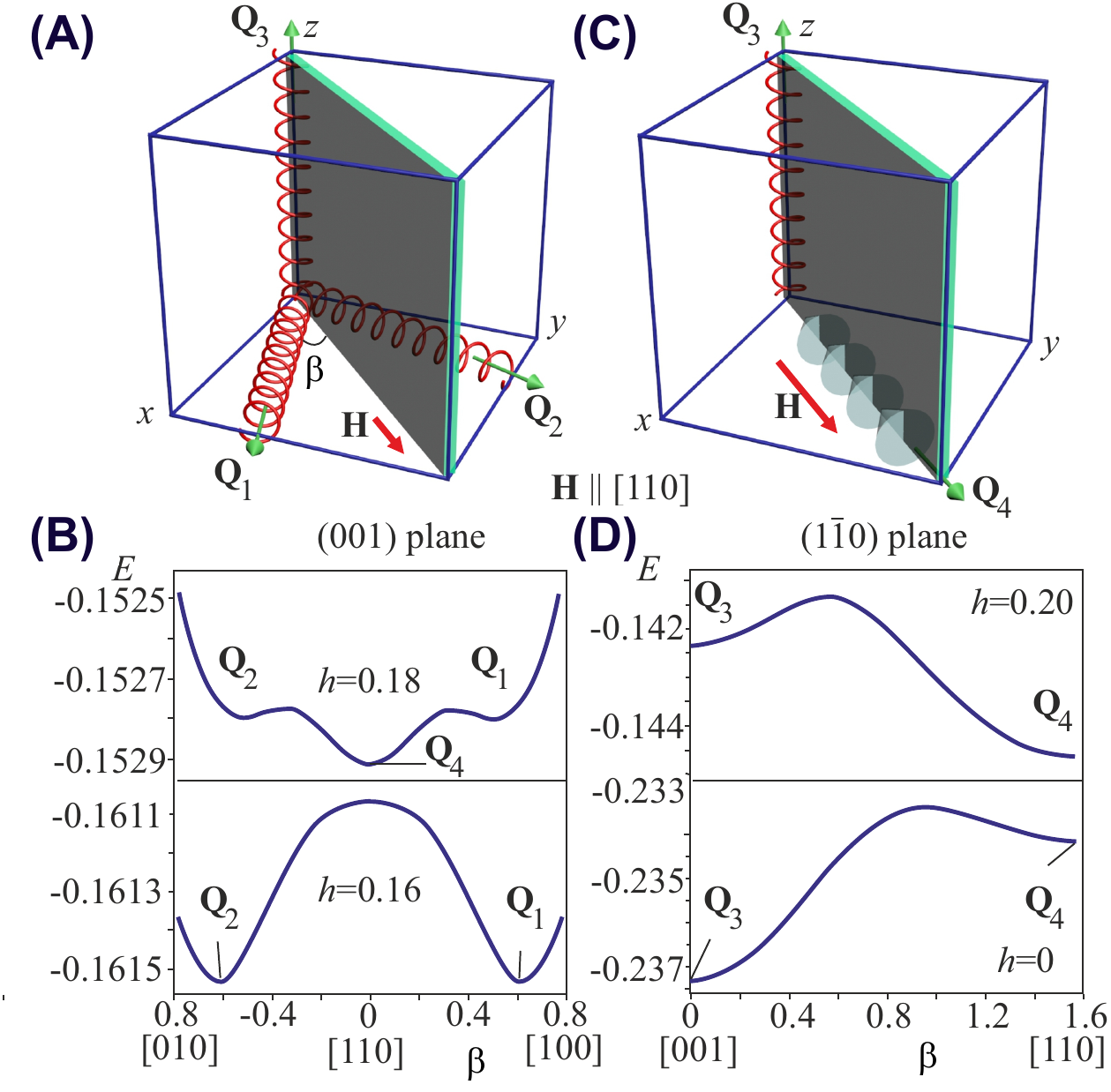}
\caption{\textbf{Spiral re-orientation for $\bm{H}\|[110]$}. 
({\bf A}) The three helical spiral states in low magnetic fields.
Under an applied magnetic field, the spiral states with $\bm{Q}_1$ and $\bm{Q}_2$ (along the [100] and [010] directions in zero field, respectively) undergo a first-order phase transition into the conical phase with $\bm{Q}_4 \| \bm{H}$ ({\bf C}), whereas the metastable spiral with the wave vector $\bm{Q}_3||[001]$  persists up to a higher magnetic field.
({\bf B}) Dependence of energy density on the angle $\beta$, given in radians, in between the spiral wave vector and the field direction below and above the first-order phase transition for $\bm Q$ varying in the (001) plane.
$\beta$-dependence of the energy density for $\bm Q$ varying in the $(1\overline{1}0)$ plane in zero field and at $h = 0.2$.  
\label{h110}
}
\end{figure}

\begin{figure}
\includegraphics[width=\columnwidth]{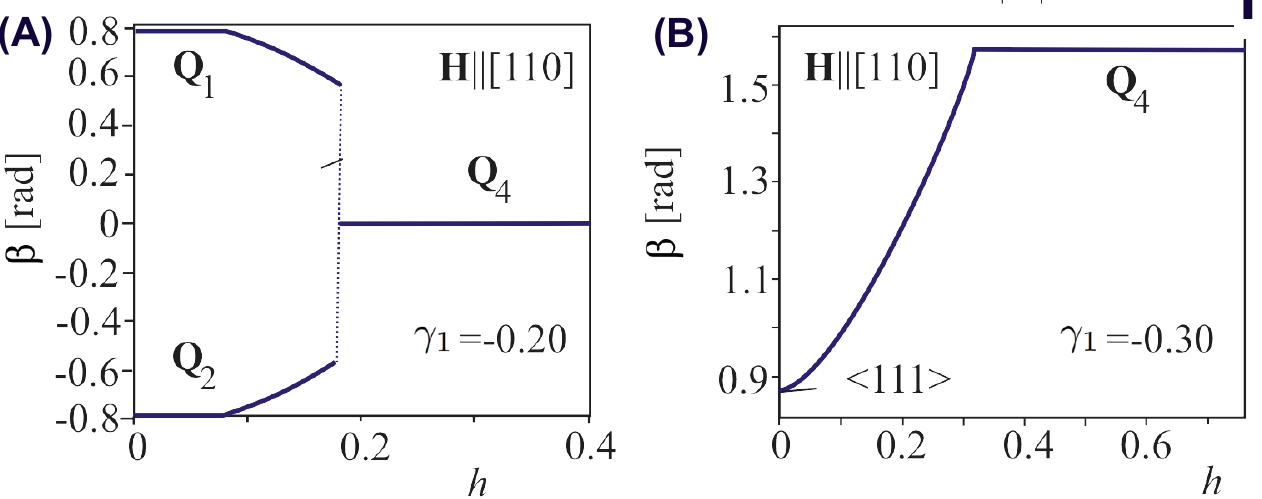}
\caption{\textbf{Field-induced re-orientation of the spiral wave vectors for $\bm{H}\|[001]$ and for the fourth-order anisotropy $\bm{\kappa}$ = -0.2 and the anisotropic exchange $\bm{\gamma_1<0}$.} 
%
({\bf A}) Field-dependence of the angle $\beta$  between  the spirals in the [001] plane and $\bm{H}\|[110]$. 
The smooth rotation of the spirals along $\bm{Q}_1$ and $\bm{Q}_2$, which at zero field are parallel to [100] and [010] respectively, is followed by a discontinuous transition into the conical spiral state with $\bm{Q}_4||\bm{H}\|[110]$ (a more detailed representation is given in Fig.~\ref{h110}). 
{\bf (B}) Smooth rotation of the spiral vector from $\langle111\rangle$ towards $\bm{H}\|[110]$ as a function of the magnetic field for a stronger exchange anisotropy, $\gamma_1 = - 0.3$. 
\label{supp_angles}
}
\end{figure}

\subsection{Effect of magnetic anisotropy on the direction of the spiral wave vector}
\label{sec:anisotropies}

For the conical spiral Ansatz Eq.(\ref{eq:Ansatz}), the anisotropy energy Eq.(\ref{eq:anisotropy}) equals
\begin{multline}
 {\cal E}_{\rm a} = 
 K \left[A(\theta) + B(\theta)(l_x^4 + l_y^4 + l_z^4)\right]
 \\  
+\frac{\sin^2(\theta)a^2}{2} \bigg[   C_1Q^2 
-C_1 \left(Q_x^2 l_x^2 + Q_y^2 l_y^2 +Q_z^2 l_z^2\right)
 \\ 
 -   
 C_2 \left(Q_z^2 l_x^2 + Q_x^2 l_y^2 +Q_y^2 l_z^2 \right) 
 + C_2 \left(Q_y^2 l_x^2 + Q_z^2 l_y^2 +Q_x^2 l_z^2\right)
 \\ 
 - 2C_3 \left(Q_x Q_yl_xl_y + Q_yQ_zl_yl_z+Q_zQ_xl_zl_x\right)
\bigg] \\
+ \frac{\sin^2\theta {\cal J} a^4}{2} \left(Q_x^4 + Q_y^4 + Q_z^4\right)
 \label{eq:anisotropy2}
 \end{multline}
with 
\begin{equation}
\left\{
\begin{array}{rcl}
A(\theta) &=& \frac{3}{8} (1 + 6\cos^2\theta - 7\cos^4\theta),\\ \\
B(\theta) &=& \frac{1}{8} (3 - 30\cos^2\theta + 35\cos^4\theta),
\end{array}
\right.
\end{equation}
where $\theta$ is the conical spiral angle.

The blue line in Fig.~\ref{fig:keff} shows the effective dimensionless anisotropy, $\kappa_{\rm eff}(\theta) = \frac{K J B(\theta)}{D^2}$, as a function of the conical angle for $\kappa = \frac{K J}{D^2} = - 0.14$. There is an interval of $\theta$, in which  $\kappa_{\rm eff}(\theta) > 0$, which favors $\bm l$ and  $\bm Q$ along a body diagonal rather than along the cubic axes.

When other magnetic anisotropies are added, $\kappa_{\rm eff}(\theta)$ is defined as the coefficient in front of $\hat{Q}_x^4 + \hat{Q}_y^4 + \hat{Q}_z^4$, where $\hat{\bm Q} = {\bm Q}/Q$. To leading order in the spin-orbit coupling constant, $\lambda$, $\kappa_{\rm eff}(\theta) =  \kappa B(\theta) - \frac{\gamma}{2} \sin^2 \theta$, where $\gamma = \gamma_1 - \gamma_3 + \gamma_4$ with  $\gamma_i = C_{i}/J$ $(i = 1,2,3)$ and $\gamma_4 = \frac{{\cal J}Q^2a^2}{J}$. The red line in Fig.~{\ref{fig:keff}} shows $\kappa_{\rm eff}(\theta)$, for $\kappa = -0.14$ and $\gamma = \gamma_1 =  -0.08$.  Negative $\gamma$ widens the interval of positive  $\kappa_{\rm eff}(\theta)$ and increases the magnitude of $\kappa_{\rm eff}(\theta)>0$ and stabilizes the tilted spiral state in some interval of magnetic field close to $H_{C2}$.


\subsection{Numerical studies of spiral re-orientation processes in presence of competing anisotropies}
\label{sec:numerics}

In this section we present more results on  the spiral re-orientation processes obtained by exact energy minimization.  We consider 
the dimensionless fourth-order and exchange anisotropies, $\kappa = \gamma_1=-0.20$, and the magnetic field $\bm{H}$ applied along the [001] and [110] directions.
Figs. \ref{h001} and \ref{h110}a, c schematically show a layout of $\bm{Q}$-vectors in stable and metastable spiral states with respect to the field and crystallographic directions, whereas Figs. \ref{h001} and \ref{h110}b,d demonstrate the energy densities of skewed spiral states depending on the $\bm{Q}$-vector direction as calculated in particular crystallographic planes.
We consider the stable orientations of $\bm{Q}$ vectors in the  (010) and $(1\overline{1}0)$ planes, Fig. \ref{h001}, and in the  (001) and $(1\overline{1}0)$ planes, Fig. \ref{h110}.

$\bm{H}\|[001]$:  
Fig.~\ref{h001} shows the three helical spiral domains with the wave vectors $\bm{Q}_1,  \bm{Q}_2$ and $\bm{Q}_3$ along the cubic axes corresponding to the global minima of the energy functional Eq.~\ref{eq:model1} in zero field.
In the applied magnetic field,  the spirals with $\bm Q_1\|[100]$ and $\bm{Q}_2\|[010]$, represented in Fig.~\ref{h001}A by red springs,  become metastable. The stable conical spiral with $\bm{Q}_3\|[001]$ is represented by blue cones. 
%
Fig. \ref{h001}B shows the dependence of spiral energy on the angle, $\phi$, between $\bm Q$  and the [001] axis, for $\bm Q$ varying in the (010) plane.
The wave vectors of the metastable spirals remain parallel to [100] and [010] these states become unstable at $h=0.32$. 
Above a critical field value, the conical spiral with $\bm{Q} = \bm{Q}_3$ begins to tilt towards one of the four body diagonals, the $[ 111]$ directions, as shown in Fig. \ref{h001}C.
Fig.~\ref{h001}D shows the dependence of spiral energy on the tilt angle for $\bm Q$ varying in the (1$\bar 1$0) plane, for $h = 0.4$ and $h = 0.5$. 
As $h$ increases, $\phi$ grows, reaches its maximal value and then decreases back to zero, corresponding to a return into the conical spiral state.  
The maximal tilt angle, $\phi_{\rm max}$, which  depends on the ratio of the competing fourth-order and exchange anisotropies, is plotted in Fig.~\ref{angles} of the main text.

$\bm{H}\|[110]$: 
The zero-field degeneracy of the helical spiral state is lifted in a different way when  $\bm{H}\|[110]$. 
Fig.~\ref{h110}A shows the three spiral domains with the wave vectors $\bm{Q}_1,  \bm{Q}_2$ and $\bm{Q}_3$ in low magnetic fields.
The magnetic field  favors the wave vectors $\bm{Q}_1$ and  $\bm{Q}_2$ in the (001)-plane. 
Fig.~\ref{h110}B shows the energy density at $h = 0.16$ and $h = 0.18$ as a function of $\beta$, the angle between $\bm{Q}_1$ and the field direction. 
The first-order phase transition between the helical spiral states with the wave vectors $\bm{Q}_1$ and  $\bm{Q}_2$ and the conical   spiral state with the wave vector $\bm{Q}_4\|\bm{H}$ (Fig.~\ref{h110}C) occurs  at $h=0.176$. 
The smooth field evolution of $\beta$ and the subsequent jump into the state with $\beta=0$, corresponding to the first-order phase transition,  are shown in Fig.~\ref{supp_angles}A of the main text. 
Under an applied magnetic field, the helical spiral state with $\bm{Q}_3||[001]$ is metastable and its wave vector is field-independent.  
It becomes unstable at $h=0.34$ (Fig. \ref{h110}D). 
The disappearance of the helical spiral states first with  $\bm{Q} = \bm{Q}_{1,2}$  and then with $\bm{Q}  = \bm{Q}_{3}$  explains the two distinct transition lines in the experimental phase diagrams (Fig.~\ref{Fig:BT}A-C) marked by  $H_{C1}^{(1)}$ and $H_{C1}^{(2)}$. 

\newpage
\subsection{Electric polarization}
\label{sec:polarisation}

The electric polarization, $\bm P$, induced by a magnetic ordering in \cuse\ is given by:
\cite{mochizuki2015dynamical}
\begin{equation}
\bm{P} = (P_a,P_b,P_c) = \lambda(m_b m_c, m_c m_a, m_a m_b),
\label{eq:P}
\end{equation}
where $\lambda$ is the magneto-electric coupling and $a$, $b$ and $c$ label the three cubic axes. Using Eq.(\ref{eq:Ansatz}) and averaging over the period of the spiral, we obtain
\begin{equation}
\langle m_i m_j \rangle = \frac{1}{2} 
\left[\sin^2 \theta \delta_{i,j} + (3 \cos^2 \theta - 1) l_i l_j \right],
\label{eq:mimj}
\end{equation}

\noindent
where $\bm{l} = \bm{e}_3$ is the spin rotation axis, $\theta$ is the conical angle and $i,j = a,b,c$. From Eqs. (\ref{eq:P}) and (\ref{eq:mimj}) one finds that 
the conical spiral state with $\bm{l} = \hat{\bm{c}} = [001]$ does not induce an electric polarization. 
Next we consider the tilted spiral state with $\bm{Q} = (\frac{1}{\sqrt{2}}\sin \phi, \frac{1}{\sqrt{2}}\sin \phi, \cos \phi)$, where $\phi$ is the tilt angle (here we chose one of the four domains, in which $\bm{Q}$ tilts towards the [111] axis). 
In the model with the fourth-order and exchange anisotropies the spin rotation axis is:
\begin{equation}
\bm{l} = (\frac{1}{\sqrt{2}}\sin \alpha, \frac{1}{\sqrt{2}}\sin \alpha, \cos \alpha),
\end{equation}
where $\alpha$ is related to $\phi$ by:
\begin{equation}
\tan \phi = \tan \alpha \frac{(1 + \gamma_1 \sin^2 \alpha)}{\left(1 + \gamma_1(1 - \frac{1}{2}\sin^2\alpha)\right)}.
\end{equation}
The average electric polarization induced by the tilted spiral is then given by Eq.(\ref{eq:Pav}).

\clearpage

\end{document}